\documentclass[12pt,a4paper]{article}

\usepackage{epsfig}
\usepackage{graphicx}
\usepackage{amsmath}

\def\nostrocostrutto#1\over#2{\mathrel{\mathop{\kern 0pt \rlap 
  {\raise.2ex\hbox{$#1$}}}
  \lower.9ex\hbox{\kern-.190em $#2$}}}
\def\lsim{\nostrocostrutto < \over \sim}   
\def\gsim{\nostrocostrutto > \over \sim}   
\def\Journal#1#2#3#4{{#1} {\bf #2}, #3 (#4)}

\def\APPB{{\em Acta Phys. Polon.} B}

\def\CPC{{\em Comp. Phys. Comm.} }
\def\EPJC{{\em Eur. Phys. J.} C}  
\def\IJMPA{{\em Int. J. Mod. Phys.} A}

\def\JETPL{{\em  JETP Lett. }}

\def\MPLA{{\em Mod. Phys. Lett.} A}

\def\NPB{{\em Nucl. Phys.} B}
\def\PLB{{\em Phys. Lett.} B}
\def\PLBB{ B}
  
\def\PRL{{\em Phys. Rev. Lett. }}
\def\PRD{{\em Phys. Rev.} D}

\def\PRpC{{\em Phys. Rep.} C}
\def\PRp{{\em Phys. Rep.} } 

\def\SJNP{{\em Sov. J. Nucl. Phys.} }
\def\SPJETP{{\em Sov. Phys. JETP} }

\def\ZPC{{\em Z. Phys.} C}
\def\YF{{\em Yad. Fiz.} }

\newcommand{\Ng}{N_g}   

\catcode`@=11
\newcount\@tempcntc
\def\@citex[#1]#2{\if@filesw\immediate\write\@auxout{\string\citation{#2}}\fi
  \@tempcnta\z@\@tempcntb\m@ne\def\@citea{}\@cite{\@for\@citeb:=#2\do
    {\@ifundefined
       {b@\@citeb}{\@citeo\@tempcntb\m@ne\@citea\def\@citea{,}{\bf ?}\@warning
       {Citation `\@citeb' on page \thepage \space undefined}}%
    {\setbox\z@\hbox{\global\@tempcntc0\csname b@\@citeb\endcsname\relax}%
     \ifnum\@tempcntc=\z@ \@citeo\@tempcntb\m@ne
       \@citea\def\@citea{,}\hbox{\csname b@\@citeb\endcsname}%
     \else
      \advance\@tempcntb\@ne
      \ifnum\@tempcntb=\@tempcntc
      \else\advance\@tempcntb\m@ne\@citeo
      \@tempcnta\@tempcntc\@tempcntb\@tempcntc\fi\fi}}\@citeo}{#1}}
\def\@citeo{\ifnum\@tempcnta>\@tempcntb\else\@citea\def\@citea{,}%
  \ifnum\@tempcnta=\@tempcntb\the\@tempcnta\else
   {\advance\@tempcnta\@ne\ifnum\@tempcnta=\@tempcntb \else \def\@citea{--}\fi
    \advance\@tempcnta\m@ne\the\@tempcnta\@citea\the\@tempcntb}\fi\fi}
\catcode`@=12

\begin{document}

\setcounter{page}{0}
\thispagestyle{empty}

\title{
QCD explanation of oscillating hadron and jet multiplicity moments 
\\  
\vspace{2cm}}

\author{Matthew A. Buican, 
\ Clemens F\"orster 
\ and  \ Wolfgang Ochs 
}

\date{
{\normalsize\it Max-Planck-Institut 
f\"ur Physik, (Werner-Heisenberg-Institut) \\
F\"ohringer Ring 6, D-80805 M\"unchen, Germany\\
}}

\maketitle

\thispagestyle{empty}

\begin{abstract}
Ratios of multiplicity moments, $H_q$ (cumulant over
factorial moments $K_q/F_q$), have been observed to show an oscillatory
behaviour with respect to order, $q$. Recent studies of $e^+e^-$ annihilations at LEP
have shown, moreover, that the amplitude and oscillation length vary
strongly with the jet resolution parameter $y_{cut}$. We study the predictions
of the perturbative QCD parton cascade assuming low non-perturbative 
cut-off ($Q_0\sim \Lambda_{QCD}\sim $ few 100 MeV) and derive the
expectations as a function of the
 $cms$ energy and jet resolution from
threshold to very high
energies. We consider numerical solutions of the evolution equations of 
gluodynamics in 
Double Logarithmic and Modified Leading Logarithmic Approximations (DLA,
MLLA), 
 as well as results from a parton MC with readjusted parameters. The main characteristics are
obtained in MLLA, while a more numerically accurate description is obtained by
the MC model. A unified description of correlations between hadrons and
correlations between jets emerges, in particular for 
the transition region of small $y_{cut}$.
\end{abstract}

\vspace{-18cm}~


\newpage

\section{Introduction}
The multiplicity of hadrons is the simplest global characteristic
of the final state in particle collisions. It was an early conjecture that
the main trends of the mean multiplicity and higher 
multiplicity moments could be reproduced by the corresponding perturbative
QCD calculations for the quark-gluon cascade. 
First results have been derived for the asymptotic high
energy limit in application of the Double
Logarithmic Approximation (DLA) \cite{fur1,bcm1,dfk1}. Taking into account the
leading contributions from the collinear and soft radiation singularities
and angular ordering \cite{ef,ahm2} 
one arrives at a probabilistic description of the
parton cascade which evolves from the high energy scale $Q$ down to the
hadronization scale $Q_0$. The perturbative expansion in the coupling
$\alpha_s$ can be resummed and the exponent can be expanded into a power
series of $\sqrt{\alpha_s}$. In this way results on mean multiplicity
\cite{ahm2,dfk1,bcmm}  and higher moments \cite{bcm1,dfk1,dfk2}
have been obtained, subsequently the next-to-leading log corrections
in MLLA \cite{bw1,lphd,ahm1,mw} and yet higher order terms 
\cite{gm,dreminosc,cdgnt}  have been derived. All these results can be obtained as
different approximations to the MLLA evolution equations for the generating
function of quark and gluon jets \cite{dkt4}. A full solution of these
equations can be obtained by numerical methods \cite{lo,lu}.

One motivation of such studies was -- and still is -- to find out how far
one can extend perturbative calculations into the regime of real 
strong interactions which determine multiparticle production.
In this way, we hope to learn more about the colour confinement mechanism.
The MLLA results on mean multiplicities and single particle spectra 
give a surprisingly good description of the data which led to the notion of 
``Local Parton Hadron Duality'' (LPHD~\cite{lphd}). The results 
here depend on
only two parameters, the QCD scale $\Lambda$ and a single non-perturbative
parameter, the transverse momentum cut-off $Q_0$ with $Q_0\gsim \Lambda$;
 an overall normalization
factor $K$ is also allowed for. 
Such parton level calculations have been applied
to a large variety of inclusive observables, in particular to
multiparticle correlations (``Generalized LPHD'' \cite{gvh});
even quasi-exclusive processes are within reach \cite{os}. In general, one observes that with increasing
accuracy, the perturbatively calculated observables are in better agreement 
with the data. 
This simple model does not contain any hadronic resonances, so it
is clear that it can only describe sufficiently
inclusive observables and should not be considered for too restricted
regions of phase space (for pertinent reviews, see
\cite{dkmt2,ko,dg,dok03}).

Whereas there is no problem with inclusive single particle results, 
the status of higher multiplicity
moments (factorial moments $F_q$ and cumulant moments $K_q$, see below), 
which are integrals over the respective  $q$ particle correlation functions,
is more controversial. 
The data on quark and gluon jets for $q\leq 5$  
\cite{opalmom}
deviate with increasing order $q$ from the higher order
logarithmic approximations~\cite{dreminasy}; they correspond to an 
expansion in $q\sqrt{\alpha_s}$ rather than $\sqrt{\alpha_s}$ \cite{dkmt2,dg} 
and the known terms fit the mean multiplicity
(moment  $q=1$) but  the moments of higher order become much larger than the
data. On the other hand, numerical solutions
of the MLLA evolution equation yield very good agreement
with these data \cite{lo,lu}; remarkably, a common fit of hadron and jet multiplicities
is possible and results in
\begin{equation}
K_{tot}\approx 1
\label{K=1}
\end{equation}
for a small cut-off $Q_0\gsim \Lambda$
where $K_{tot}$ is the ratio of the total hadron to parton multiplicities.
 This ratio $K_{tot}\approx 1$ was also found to be consistent with
the quark and gluon multiplicities in high $p_T$ jets observed at the
TEVATRON \cite{cdf} where the higher order MLLA corrections are taken into
account as well.

An interesting prediction \cite{dreminosc} 
from the asymptotic $\sqrt{\alpha_s}$ expansion concerns the ratio of moments
\begin{equation}
H_q= K_q/F_q 
\label{hqdef}
\end{equation}  
These moments show an oscillatory behaviour with the first minimum at
\begin{equation}
q_{min}\approx \frac{1}{h_1\gamma_0}+\frac{1}{2} +{\cal O}(\gamma_0),\quad 
  h_1=\frac{11}{24}, \quad \gamma_0^2=\frac{2N_C\alpha_s}{\pi}
\label{qmin}
\end{equation}  
where $N_C=3$ and $\gamma_0$ denotes the leading order multiplicity
anomalous dimension at the considered energy scale, numerically $q_{min}\approx 5(\pm1)$ at LEP energies.
Such oscillations have indeed been observed in
$e^+e^-$ annihilations at SLC \cite{slacosc} and recently at LEP
\cite{L3osc,mangeol} with the first minimum at  $q_{min}\approx 5$ as expected.

Since the absolute size of the moments are not in quantitative agreement 
with these high energy predictions
the physical origin of these oscillations remains controversial.
It has been noted that 
the truncation of the multiplicity distribution at large 
multiplicity, $n$, also gives rise to
oscillations~\cite{glu1}. Furthermore, a superposition of 
two components (like 2
and 3 jet events), each one modeled without oscillations, 
can lead to oscillations
in the full sample \cite{glu2}.

A new element  has been brought into the discussion recently 
through analyses of the L3 measurements \cite{L3osc,mangeol}
of multiplicity moments for hadrons and
narrow jets at high resolution (small $y_{cut}$ parameter).
The measurements feature a rapid variation of both the 
oscillation amplitude and
oscillation length in $q$ as function of $y_{cut}$. 
This behavior does not appear in
the asymptotic solutions of the evolution equation \cite{dreminosc}.

In this paper we aim at an understanding, within perturbative QCD, of the $H_q$ oscillation
phenomena in both hadron
and jet final states.
We study first the solutions of the evolution equations
(both DLA and MLLA) 
with jet hardness from threshold  up to the asymptotic regime; furthermore,
we investigate the variations of the jet multiplicity moments with resolution, from  
low resolution (fat jets) up to high resolution (narrow jets) -- 
ultimately up to the
scale for the parton-hadron transition.
These calculations depend only on the QCD scale,
$\Lambda$, and on one parameter, $Q_c$, which refers to the
transverse momentum cut-off. This transverse momentum cut-off is given by either the scale of the 
arbitrarily chosen jet resolution 
or the characteristic hadronic scale
$Q_c=Q_0$.
The full results including the dependence on $Q_c$, 
are obtained by numerical methods.
The need for an overall normalization factor $K$ is considered for
the mean multiplicity. 

A similar theoretical scheme is realized in a
parton level MC (ARIADNE \cite{ARIADNE}) which is based on  
sequential parton radiation from colour
dipoles \cite{gus,adkt} with a $k_T$ cut-off. We have readjusted the parameters 
$\Lambda$ and $Q_0$ in this MC 
to describe hadronic final states without an additional
hadronization phase, assuming again a duality between hadron and parton 
final states at scale $Q_0$.

\section{Definition of moments and jet resolution}
The distribution, $P_n$, of the multiplicity, $n$, of particles or jets in an
event can be characterized by its moments. One considers the factorial
moments $f_q$ or the normalized moments $F_q$
\begin{equation}
f_q=\sum_{n=0}^\infty n(n-1)\ldots (n-q+1)P_n, 
   \quad  F_q=f_q/N^q, \quad  N\equiv f_1
\label{fmom}
\end{equation}
with mean multiplicity $N$. Furthermore, one introduces the cumulant moments 
$k_q$ and $K_q$ which are used to measure the genuine correlations without 
 uncorrelated background in a multiparticle sample
\begin{equation}
k_q=f_q-\sum_{i=1}^{q-1} {q-1 \choose i} k_{q-i} f_i, \qquad K_q=k_q/N^q,
\label{kmom}
\end{equation}
in particular $K_2=F_2-1,\ K_3=F_3-3F_2+2$; for a Poisson distribution
$K_1=1,\ K_q=0$ for $q>1$.

These moments can also be obtained from the generating function of the
multiplicity distribution
\begin{equation}
Z(Y,u)=\sum_{k=0}^\infty P_n(Y) \ u^k
\label{genf}
\end{equation}

by differentiation
\begin{equation}
f_q=\frac{d^q\ }{du^q} Z(Y,u)|_{u=1}, 
   \qquad  k_q=\frac{d^q\ }{du^q} \ln Z(Y,u)|_{u=1},
\label{fkgen}
\end{equation}
where $Y$ denotes here some kinematic variable like the total energy.

In a high energy collision, many hadrons are produced
and they are found to cluster typically into jets of particles.
The number of such jets depends on the resolution,
which is defined through
an algorithm in terms of a resolution parameter. In this paper, we choose the
so-called Durham algorithm \cite{durham}. 
At resolution $y_{cut}$ one 
combines particles into jets iteratively
until all pairs of jets or particles have relative transverse momentum
$k_T^2>y_{cut}Q^2$ for small relative angles $\Theta_{ij}$, or, in general, 
\begin{equation}
y_{ij}=2\min(E_i^2,E_j^2)(1-\cos\Theta_{ij})/Q^2\ > \ y_{cut}.
\label{durham}
\end{equation}
In our application to $e^+e^-$ annihilation, $Q$ is the total $cms$ energy.
At very low resolution ($y_{cut}\to 1$)
one sees only two jets which evolve from the primary $q\overline q$ system, 
whereas with increasing
resolution more and more jets are resolved until at very high resolution
($y_{cut}\to 0$) all final state hadrons are resolved, so 
\begin{equation}                                          
y_{cut}\to 1: \quad N_{jet}=2, \qquad\qquad  
y_{cut}\to 0: \quad N_{jet}=N_{hadron}.
\label{ycutlimits}
\end{equation}
In this paper we study both the multiplicity of jets and the multiplicity of
 hadrons and the respective moments. In particular, we are interested in the
variation of these observables with energy and resolution as has recently
been reported in the experimental works \cite{L3osc,mangeol}. 

In the popular Monte Carlo models of today, the parton cascade is terminated
at a scale of $Q_0\sim 1$ GeV; this perturbative phase is followed by the
hadronization phase for which non-perturbative models, typically with a
rather large number of parameters, are introduced. In our calculation we
follow the idea of LPHD \cite{lphd}
and let the parton cascade evolve further down to a lower scale $Q_0$
of a few 100 MeV of the order of the QCD scale $\Lambda$ itself. Then we
compare the result directly to the hadron final state. In this way, besides
the common parameter $\Lambda$ only one extra non-perturbative parameter 
$Q_0$ is
introduced which can be interpreted either as a typical hadron mass or as
inverse hadron radius. In the theoretical calculations we obtain the
hadronic final state for $Q_c\to Q_0$ and we replace (\ref{ycutlimits}) by
\begin{equation}                                          
y_{cut}\to 1: \quad N_{jet}=2, \qquad\qquad  
y_{cut}\to (Q_0/Q)^2: \quad N_{jet}=N_{hadron}.
\label{ycutlimitshad}
\end{equation}

 An important feature is the transition from jets to hadrons for
small resolution parameter $y_{cut}$ which tests the hadronization models
in the region of small momentum transfers, the so-called ``soft'' region.
In this paper the consequences of such an approach for multiparticle
correlations in the soft region is investigated in the theoretical models
and compared with the data. 

\section{Solution of evolution equations for multiplicity moments}
\subsection{Double Logarithmic Approximation}

In this approximation we consider a single jet of
particles or sub-jets 
at transverse momentum cut-off $k_T>Q_c$; for $e^+e^-$ annihilation, two jets
have to be superimposed, with $Q_c$ then corresponding to the resolution $y_{cut}$. 
Only the most singular terms of the parton splitting functions are kept
and recoil effects are neglected.  
Then we can restrict ourselves to gluodynamics, i.e. we neglect the production
of quark anti-quark pairs which would represent a nonleading contribution.
A common simplification
of these calculation is the restriction to 1-loop results; furthermore, in
this application we consider only light quarks.
In this case, the generating function
follows an evolution equation \cite{dfk1,dfk2,bcm} 
\begin{align}
\frac{dZ(Y_c,u)}{dY_c}& 
        =Z(Y_c,u) \int_0^{Y_c} dy' \gamma_{0}^2(y')[Z(y',u)-1]
         \label{dlaevol}\\
Z(0,u) &=u \label{init}
\end{align}
with the multiplicity anomalous dimension
\begin{align}
  \gamma_{0}^2(y) & = \frac{2N_C\alpha_s(y)}{\pi} = 
           \frac{\beta^2}{y+\lambda_c}, 
     \label{gamma0}\\
Y_c =\ln(E/Q_c),\quad \lambda_c & =\ln(Q_c/\Lambda), \quad
 \beta^2=\frac{4 N_C}{b},
   \quad b=\frac{11}{3}N_C-\frac{2}{3} n_f
     \label{gamma0par}
\end{align}
where $N_c=3$. The argument of $\alpha_s$ is related to the transverse momentum, i.e.
 $y=\ln(k_T/Q_c)$. In our analytic calculations we take 
\begin{equation}
\Lambda=300\ \text{MeV},\quad \lambda=\ln (Q_0/\Lambda)=0.015,\quad n_f=3.
\end{equation}
The initial condition (\ref{init}) implies that at threshold 
($Y_c=0$) the jet $Z(Y_c,u)$ contains only one particle (sub-jet).

Differentiating the evolution equation (\ref{dlaevol}) 
with respect to $u$ at $u=1$, one obtains the following
equation for the multiplicity, $N$:
\begin{equation}
N'(Y_c)=\int_0^{Y_c} dy' \gamma_{0}^2(y') N(y')
\label{nprimedla}
\end{equation}
or, equivalently,
\begin{equation}
N''(Y_c) -\gamma_0^2(Y_c)N(Y_c) =0, \qquad N(0)=1,\quad N'(0)=0.\label{neqn}
\end{equation}
This equation can be solved in terms of Bessel functions \cite{dfk1}
\begin{equation}                             
N(Y_c)=2\beta \sqrt{Y_c+\lambda} 
\left[ I_1(2\beta\sqrt{Y_c+\lambda}) K_0(2\beta\sqrt{\lambda}) +
K_1(2\beta\sqrt{Y_c+\lambda}) I_0(2\beta\sqrt{\lambda}) 
\right]
\label{dlamult}
\end{equation}

\begin{figure}[t!]
\begin{center}
\includegraphics[angle=-90,width=14cm]{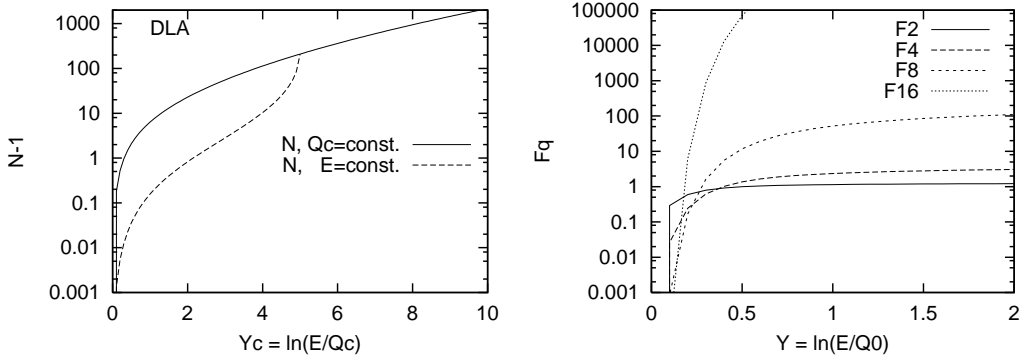}
\end{center}
\vspace{-5.0cm}
\caption{
Multiplicity, $N$, in DLA in a single jet
 vs jet energy $E$ for fixed $Q_c=Q_0$ (representing hadrons)
 and for fixed energy $E$ ($Y_0=\ln(E/Q_0)=5$)
but variable resolution $Q_c$ (or $y_{cut}=(Q_c/2E)^2$) (representing
jets) and
factorial moments $F_q$ vs. energy $E$ ($Y$). The moments approach the asymptotic limits $F_2=1.33, F_4=4.62,
F_8=359.7, F_{16}=2\times10^8$.
}
\label{hqdla-nf}
\end{figure} 
This solution is shown in Fig. \ref{hqdla-nf}, for constant cut-off
$Q_c=Q_0$ as a full line, which represents the well known rise of
multiplicity, 
$N\sim \exp(2\beta\sqrt{E/\Lambda})$, and follows from the asymptotics
$I_\nu(z)\sim e^z$. 

One can also study the variation of jet multiplicity
at fixed jet energy $E$ with resolution $Q_c$, as discussed previously
\cite{lo}. This dependence is given by the same
function (\ref{dlamult}) if the Durham algorithm is applied; in this case
the cut-off in the evolution of the parton cascade is defined through the
transverse momentum $k_T\geq Q_c$. The corresponding result is
represented by the dashed line in
Fig. \ref{hqdla-nf}. One observes a considerably lower multiplicity
by an order of magnitude, but, for decreasing $Q_c \to Q_0$ ( $Y_c\to Y_0$)
in the transition from jet $\to $ hadron, the jet curve rises rapidly and reaches the
curve for hadrons. This behaviour is a consequence of the running coupling,
which is always smaller for jets, because $k_T$, the argument of $\alpha_s$, is
always larger. The analytic behaviour in this limit can be derived from
(\ref{dlamult}) using the approximation $K_0(z)\simeq \ln(2/z)$ for small $z$
with small $Q_c$ or large $Y_c$ (see also \cite{ko})
\begin{equation}
    N\sim
(\beta^2Y_c)^{1/4} \ln\left(\frac{1}{\beta\sqrt{\lambda_c}}\right)
     \exp{2\beta \sqrt{ Y_c}}. 
\label{resolvedla}  
\end{equation}
This multiplicity diverges logarithmically 
for $Q_c\to \Lambda$ ($\lambda_c\to 0$). This divergence is a consequence of the diverging coupling in this limit.
Since $Q_c\geq Q_0>\Lambda$, this divergence is outside the physical region, but a strong variation remains,
as seen in Fig. \ref{hqdla-nf}.
Note that in the case of fixed coupling, the moments would only depend on the
ratio $E/Q_c$ as no other dimensional quantities exist; then both curves
for jets and hadrons would coincide  and follow asymptotically a power law
$N\sim (E/Q_0)^{\gamma_0}$.

Next we calculate the factorial moments for $q>1$. By 
appropriate differentiation
of the evolution equation (\ref{dlaevol}) one finds for  $q>1$:
\begin{eqnarray}
f_q'(Y_c)&=& \int_0^{Y_c} \hspace{-0.2cm} 
                      dy' \gamma_0^2(y') f_q(y') 
         + \sum_{m=1}^{q-1} {q \choose n} f_m(y) \int _0^{Y_c} 
             \hspace{-0.2cm} dy' 
              \gamma_0(y') f_{q-m}(y') \\
f_q(0)&=&0.
\label{fmomeq}  
\end{eqnarray}
where (\ref{fmomeq}) follows from (\ref{init}) and (\ref{fkgen}),
in agreement with (\ref{fmom}).

This coupled system of 
integro-differential equations is solved numerically. The integral
is computed using the trapezoidal rule with a step size of $\delta
y=10^{-5}$. 
As a test, we calculated the
multiplicity from the analytic formula and compared with the numerical
solution of (\ref{nprimedla}). 
There was good agreement to within $10^{-4}$. Reducing the step size by a factor of 10 ($\delta y = 
10^{-6}$), resulted in a change of about 1\% in the value of $F_{16}$ and a change in the ratio
$H_5$ by 0.1\% below in the range considered.

A selection of results on the normalized moments $F_q$ 
is also shown in Fig.~\ref{hqdla-nf}. 
The moments $F_q$ for $q>1$ vanish at threshold
$Y_c=0$. Shortly above threshold the moments of higher rank $q$
are suppressed as the multiplicity is low here 
and only a small
number of terms contribute to the sum (\ref{fmom}); on the other hand, they
 approach larger asymptotic values. In DLA one finds \cite{bcm1}
\begin{equation}
F_q=\frac{q}{q^2-1}\sum_{m=1}^{q-1} 
 {q \choose m} \frac{F_mF_{q-m}}{m}\quad \text{for}\ q>1\quad
   \text{with}\quad  F_1 = 1,
\label{fqdla}
\end{equation}
i.e. for the first terms
\begin{equation}
 F_2=\frac{4}{3},\quad  F_3 = \frac{9}{4},\quad 
 F_4 = \frac{208}{45},\quad F_5 = \frac{2425}{216}, \quad 
      F_6 = \frac{2207}{70}.
 \label{f1to6}
\end{equation}
Therefore, with increasing energy more and more moments $F_q$ are needed to
represent the multiplicity distribution.
 
Alternatively, one can study the ratios $H_q=K_q/F_q$ of cumulant and
factorial moments. An intesting feature of these ratios is their 
asymptotic decrease with rank $q$ \cite{dg}
\begin{equation}
H_q \simeq 1/q^2.     \label{hqasy}
\end{equation}

\begin{figure}[t!]
\begin{center}
\includegraphics[angle=-90,width=10cm]{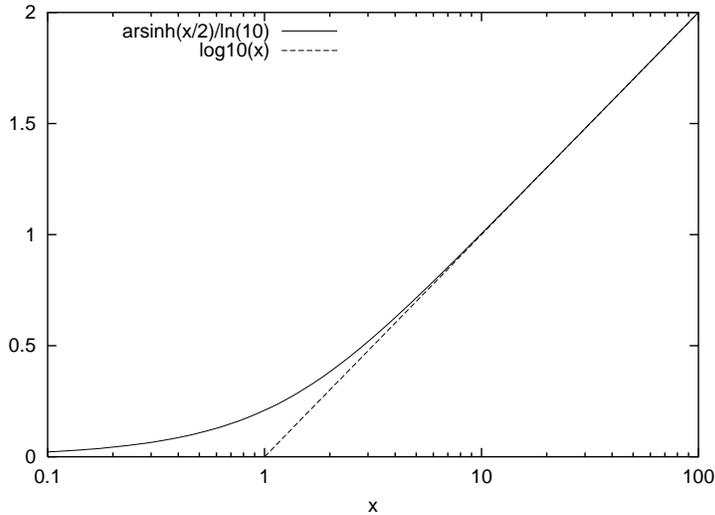}
\end{center}
\vspace{-0.3cm}
\caption{
The function $y={\rm Ash}_{10}(x)\equiv {\rm Arsinh}(x/2)/ \ln(10)$  represents data 
with positive and negative signs over a large range of scales,
in comparison with the common $y=\log_{10}x$.
}
\label{arsinh-test}
\end{figure}

\begin{figure}[t!]
\begin{center}
\includegraphics[angle=-90,width=12cm]{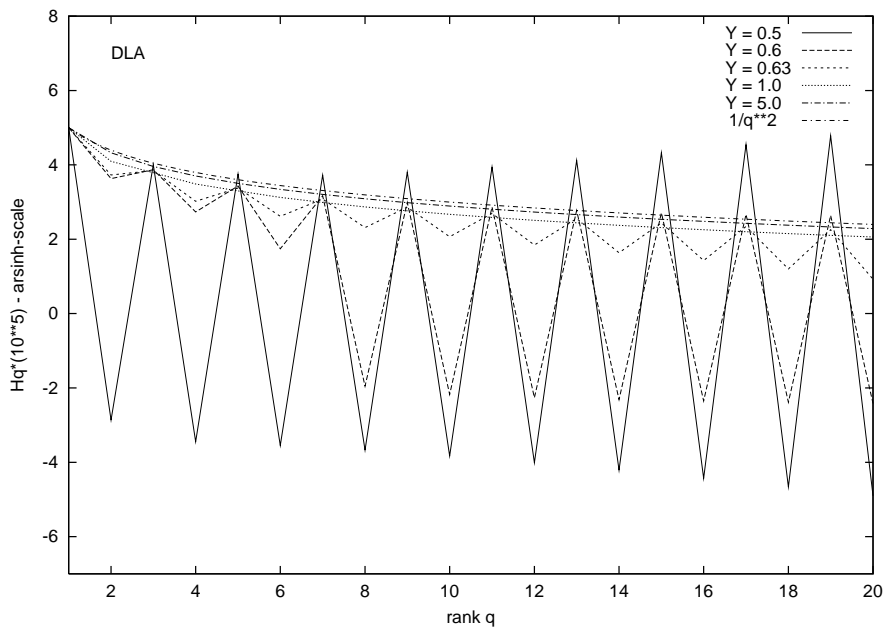}
\end{center}
\vspace{-0.3cm}
\caption{
Moment ratios $H_q$ as function of rank $q$ in DLA
for different energies $Y$
for fixed $Q_c=Q_0$ (representing hadrons).
The quantity plotted is
${\rm Ash}_{10}(H_q\times 10^{5})$ with function (\ref{ash10}).  
With increasing $Y$ the moments approach the 
asymptotic behaviour 
 $H_q \sim 1/q^2$ 
}
\label{hqdla-q}
\end{figure} 

This behaviour can be derived easily in leading order of $\gamma_0 \sim \sqrt{\alpha_s}$ under the assumption of 
constant $K_q$ and $F_q$ from the equation $k_q''=\gamma_0^2f_q$, which follows from (\ref{fkgen}) and 
(\ref{dlaevol}) using (\ref{neqn}).

On the other hand, these ratios are not very convenient near threshold.
First, we can derive for the cumulant moments 
from (\ref{kmom}) and (\ref{fmomeq})
\begin{equation}
K_q=(-1)^{q-1}(q-1)!\quad {\rm for} \ Y_c=0.   \label{kmom0}
\end{equation}
Then, because $F_q=0$ for $q>1$, the ratios $H_q$ diverge 
in the limit $Y_c\to 0$ with alternating signs. Note that the observables
$F_q,K_q,H_q$ each provide a full description of the multiplicity
distribution. At high energies $K_q$ and $H_q$ are convenient, whereas at low energies $F_q$ is
more convenient since the respective higher order terms are suppressed. 

Before we come to a discussion of the results, we introduce a convenient way
to present data with a large range of scales and alternating signs.
It is an extension of the usual log plot for positive numbers 
in which a positive quantity $y$ is
represented in log scale $y=10^x,\ x=\log_{10}y$. For a quantity with
either sign we write
\begin{eqnarray}
y=10^x - 10^{-x}= 2 {\rm Sinh}(x\ln10)\\
x:={\rm Ash}_{10}(y)=(1/\ln10) {\rm Arsinh}(y/2) \label{ash10}
\end{eqnarray}
where ${\rm Arsinh}(z)=\ln(z+\sqrt{1+z^2})$. For large positive or negative
numbers one obtains ${\rm Ash}_{10}(\pm y)\approx \pm\log_{10}(y)$
For convenience we show in Fig. \ref{arsinh-test} the comparison
of both functions.

First, we study the evolution of the ratios, $H_q$, with energy $Y=\ln(E/Q_0)$
and this is shown in Fig. \ref{hqdla-q} for a few values of $Y$.
One can see the oscillations with large amplitude near threshold (small
$Y$) which follow from (\ref{kmom0}).
With increasing energy the oscillations continue with the same
oscillation length but with decreasing amplitude. The moments of even rank
$q$ finally change sign and the oscillations  
disappear at $Y=1$ where they are already close to the asymptotic
limit (\ref{hqasy}). Also one observes the rise of $H_q$ with $q$ at small
$Y$ and the decrease of $H_q$ with q at large $Y$.

\begin{figure}[t!]
\begin{center}
\includegraphics[angle=-90,width=13cm]{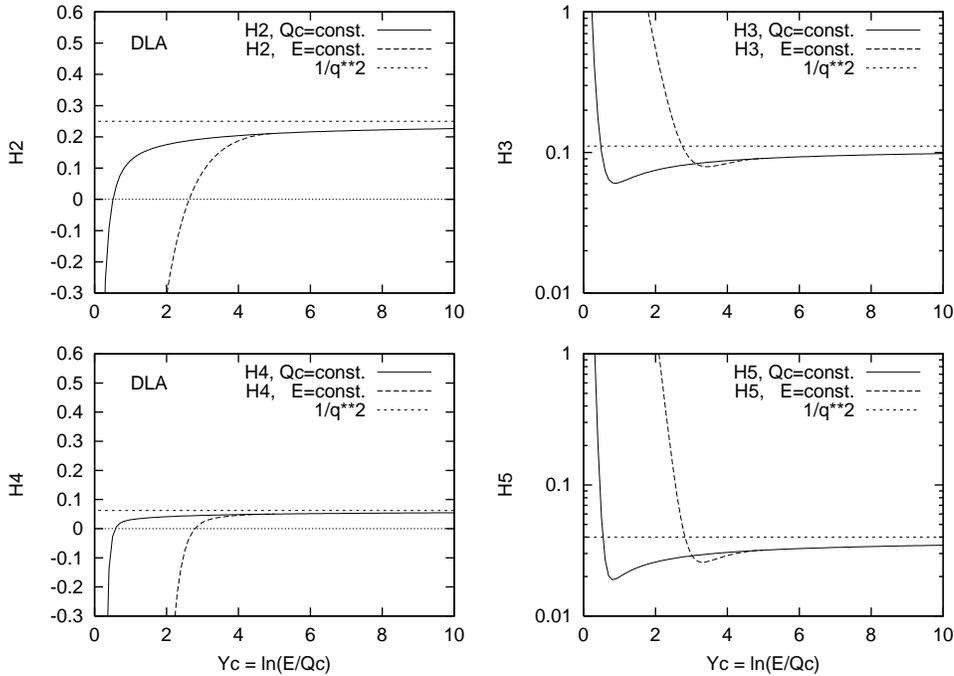}
\end{center}
\vspace{-0.3cm}
\caption{
Ratios of moments $H_q$ in DLA vs energy for fixed $Q_c=Q_0$ 
(representing hadrons) and for fixed energy $Y=\ln(E/Q_0)=5$
 but variable $Q_c$ representing
jets at variable resolution $y_{cut}=(Q_c/2E)^2$. The asymptotic limits
$H_q\to 1/q^2$ are shown as well.
}
\label{hqdla-yc}
\end{figure} 

Next we present in Fig. \ref{hqdla-yc} the evolution of the ratios $H_q$
 with energy $Y$ which diverge at threshold and approach the asymptotic
result (\ref{hqasy}) from below.
At the same time we show the variation of the multiplicity moments for
jets at variable resolution at fixed energy $Y=5$ (corresponding to LEP
energies at $Q_0\approx 0.3$ GeV). Similarly to the case of
multiplicity, there is a large difference between both dependences,
which reflects again the role 
of the running coupling. The rise of the jet moments
is delayed but they reach the hadron moments as $Q_c\to Q_0$ 
at the nominal energy of $Y=5$.

Finally we remark that the comparison with $e^+e^-$ results would 
require quark jets in two hemispheres. This can be obtained \cite{dfk1} by
replacing the generating function as
\begin{equation}
Z\ \to\  Z^{2C_F/C_A}=Z^{8/9}.
\end{equation} 
We do not go into this minor difference in the present qualitative
discussion.

\subsection{Modified Leading Logarithmic Approximation}

In this approximation, the full DGLAP splitting functions \cite{ap,dok} 
are included, with energy
conservation taken into account. The restriction to 1-loop results
remains. MLLA takes into account the next-to-leading order terms in the
$\sqrt{\alpha_s}$ expansion of the exponent, $\ln \Ng$. In the present
analysis, we neglect the $q\bar q$ pair production (even though it contributes in MLLA)
for simplicity -- it could be taken into account without
difficulty (see, for example \cite{lo,lu}).
The evolution equation for the generating function $Z(Y_c,u)$ 
in gluodynamics reads \cite{dkt4}
\begin{eqnarray}
\frac{dZ(Y_c,u)}{dY_c} &=& \int_{z_c}^{1-z_c} dz 
   \frac{\alpha_s(\tilde{k_T})}{2\pi}P_{gg}(z)\times \nonumber \\
   && \quad\ \times    \{Z(Y_c + \ln z,u)Z(Y_c + \ln (1-z),u) - Z(Y_c,u)\}
   \label{mllaevol}\\
Z(0,u)&=& u  \label{mllainit}
\end{eqnarray}
where 
\begin{equation}
z_c = e^{-Y_c} ,\quad \alpha_s(\tilde{k_T}) = \frac{2\pi}{b\,\ln
(\tilde{k_T}/\Lambda)} ,\quad \tilde{k_T} = \min (z, 1-z)E 
\end{equation}
and the splitting function $P_{gg}(z)$ is given by\footnote{The 
$P$ function \cite{ap} is related to the  $\Phi$ function
 \cite{dok} by  $\Phi_{ab}=2
P_{ab}$}
\begin{equation}
 P_{gg}(z) = 6 \left[z(1-z)+\frac{1-z}{z} + \frac{z}{1-z}\right].
   \label{phiasy}
\end{equation}
For the symmetric kernel one can replace in 
the integral $\frac{1}{2}P_{gg}(z)$ by  $P^{asy}_{gg}(z)=(1-z)P_{gg}$
which has no pole at $z = 1$ and behaves for $z\to 0$ like 
$P^{asy}_{gg}\sim 6/z$.
Differentiation by $u$ leads again to evolution equations for the
multiplicity $\Ng$ 
\begin{align}
\Ng' (Y_c) = \int_{z_c}^{1-z_c} dz 
     & \frac{\alpha_s(\tilde k_T)}{\pi} P_{gg}^{asy}(z)\times \nonumber\\
        &\times \{\Ng(Y_c+\ln z)+\Ng(Y_c+\ln (1-z))-\Ng(Y_c)\}\label{multevol}
\end{align}
and the unnormalized factorial moments
\begin{align}
f_q'(Y_c) = \int_{z_c}^{1-z_c} dz
      &\frac{\alpha_s(\tilde k_T)}{\pi} P_{gg}^{asy}(z) \times  \nonumber\\
      &\times \{f_q(Y_c+\ln z) + f_q(Y_c + \ln (1-z)) - f_q(Y_c) + \nonumber \\
 & + \sum_{m = 1}^{q-1}{q
        \choose m} f_m(Y_c+\ln z)f_{(q-m)}(Y_c + \ln (1-z))\}\label{fqevol}
\end{align}
At threshold for multiparticle production, we find the following conditions
\begin{equation}
q=1:\quad \Ng=1\ \text{for}\ E\leq 2Q_c, \qquad
q>1:\quad  f_q=0\ \text{for}\ E\leq q Q_c. \label{initmlla} 
\end{equation}
The condition for $q=1$ follows from the initial condition $N_{g}=1$ and $N_{g}'=0$
in (\ref{multevol}) below the threshold for particle emission at $E\leq 2Q_c$. 

The constraint for the 
higher moments, $f_q$ for $E<q Q_c$, corresponds to the kinematic condition
for producing $q$ particles.
We can derive the condition for $q>1$ by induction. First,
consider the case of $q=2$ in (\ref{fqevol}). We know
$f_2(Y_c)=0$ if $Y_c\leq \ln 2$ (equivalently, if $E
\leq 2Q_c$) from  $f_1\equiv N=1$ in this
region. Now let us consider the situation for $q>2$, taking
(\ref{initmlla}) for $q-1$ as given.
Consider first the evolution of $f_q(Y_c)$ in 
(\ref{fqevol}) with $Y_c$ starting
from $f_q=0$ at $Y_c\leq \ln 2$ where $N=1$. As long as 
the sum in (\ref{fqevol}) vanishes
we just have the evolution $f_q'(Y_c)=f_q(Y_c)\times I(Y_c)$ with a known
integral $I$ 
and therefore $f_q=0$.

Next we show that for energies  $E \leq qQ_c$ (or $Y_c<\ln q$)
this sum indeed vanishes. First, note that 
the argument $\ln(zE/Q_c)$ of the moment $f_m$ varies from $0$ at the lower
limit to $\ln(q-1)$ at the upper limit while $z$ varies from $z_c$ to
$1-z_c$ and the argument
of $f_{(q-m)}$  decreases from  $\ln(q-1)$ to $0$. 
In the sum, exactly one of the factors --
$f_m$ at energy $E'=zE$ or $f_{q-m}$ at energy $E''=(1-z)E$ 
-- vanishes because of condition 
(\ref{initmlla}) for $m<q$ or $q-m<q$
\begin{eqnarray}
E'<mQ_c \ \ E''>(q-m)Q_c & f_m=0 & f_{q-m} \geq 0 \\
E'>mQ_c \ \ E''<(q-m)Q_c & f_{q-m} = 0 & f_m \geq 0
\end{eqnarray}
Now, the perceptive reader may notice that for $m=1$, $f_m = \Ng$ cannot be $0$. This is
 no problem, however, since $f_{q-1}=0$ in the whole range. Thus, we have proven
the validity of (\ref{initmlla}) for any $q$.

The MLLA evolution equation (\ref{mllaevol}) reduces to the DLA equation
(\ref{dlaevol}) if the integrand is taken in small $z$ approximation,
 i.e. the $\ln (1-z)$ term is neglected and $P_{gg}(z)\sim 1/z $. This limit 
should be achieved for very high energies, because $\alpha_s(\tilde k_T)$ remains
energy independent and gives weight to ever smaller values of $z$ under the
integral. Indeed, the high energy results for the moments in terms of an
$\sqrt{\alpha_s}$ expansion \cite{mw,dg} yields the DLA solutions in the
asymptotic limit. 

Full solutions of the evolution equations (\ref{multevol}),(\ref{fqevol})
satisfying the threshold condition, analogous to (\ref{dlamult}) in DLA
are not known. If the integrand is simplified assuming a $\sqrt{\alpha_s}$
expansion one obtains expressions with Bessel functions similar to
(\ref{dlamult}) but with non-integer index \cite{lphd,cdfw} which remain
finite for $Q_0\to \Lambda$. 

Here we solve the system of equations (\ref{multevol}),(\ref{fqevol}) 
again numerically using the trapezoidal
approximation with step size $10^{-3}$ 
for the integral  (for control also with $10^{-2}$),
alternatively the 3-point Simpson rule has been applied with
similar results. Note that the
evolution variable $Y_c\sim \ln E$ now not only appears in the integral bound
but also in the integrand which was not the case in the DLA.  

\begin{figure}[t!]
\begin{center}
\includegraphics[angle=-90,width=14cm]{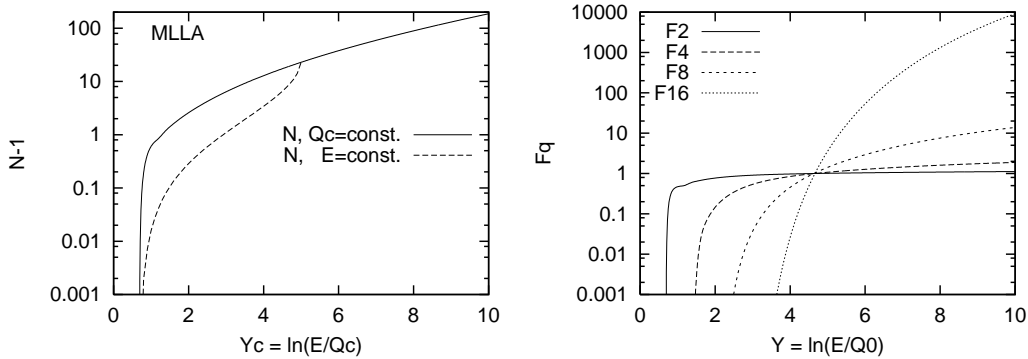}
\end{center}
\vspace{-5.0cm}
\caption{
Multiplicity $N$ in DLA in single jet
 vs energy for fixed $Q_c=Q_0$ 
(representing hadrons) and for fixed energy $Y_0=\ln(E/Q_0)=5$
 but variable $Q_c$ representing
jets at variable resolution $Q_c$ (or $y_{cut}=(Q_c/2E)^2$) and
factorial moments $F_q$ vs. energy $Y_0$. 
}
\label{hqmlla-nf}
\end{figure} 

In Fig. \ref{hqmlla-nf} we show the numerical results for the multiplicity 
$N$ in a single jet as function of $Y_c$ for hadrons (fixed  $Q_c=Q_0$ 
according to LPHD) and for jets at fixed energy $E$ (LEP energy) but
variable resolution. The results look similar to Fig. \ref{hqdla-nf} in DLA,
but the absolute size is now much reduced because of the energy conservation
constraint. 

Also shown are some factorial moments, $F_q$. Again, the moments are smaller
in size and, therefore, the approach to the asymptotic values (\ref{f1to6}) 
is further delayed. A distinctive difference from DLA are the shifted threshold
energies for the higher moments according to (\ref{initmlla}).

The factorial moments $F_q$ show a striking result at intermediate energies: 
they all cross
at one point ($E \approx 20$ GeV) with $F_q=1$, i.e. with a Poissonian
distribution
\begin{equation}
\text{Poissonian transition point:} \quad F_q=1 \qquad \text{at} 
\qquad Y_P \approx 4.7.   \label{poissonpt}
\end{equation}
This feature does not appear in DLA, and is apparently related to the delayed
thresholds. For the considered moments with $q\leq 16$ 
the crossing at $F_q=1$  occurs within an
interval of $\delta Y\approx 0.05$ for both step sizes $10^{-3}$ and
$10^{-2}$ of our numerical computation, so the distribution is close 
to a Poissonian. We note though, that the higher moments
 $F_q$ with  $q\gsim 32$ have a threshold above
the transition point $Y_P$ so the Poissonian cannot be an exact solution.
Nevertheless, it is remarkable that besides the threshold and asymptotic
regime, there is an intermediate energy with a very simple behaviour. 
The Poissonian transition
point follows apparently from the evolution equation -- a novel property 
for which we have no analytical explanation yet.

\begin{figure}[t!]
\begin{center}
\includegraphics[angle=-90,width=12cm]{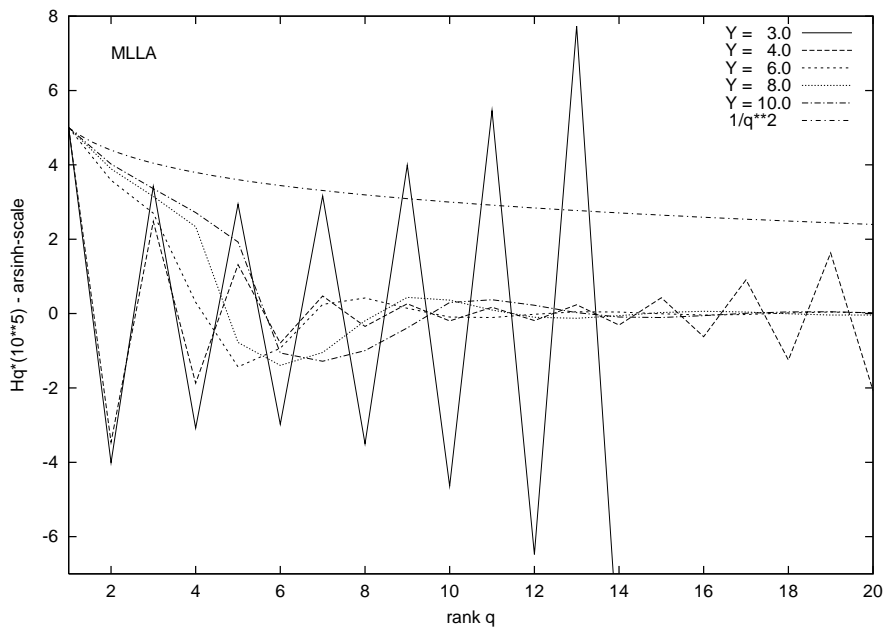}
\end{center}
\vspace{-0.3cm}
\caption{
Moment ratios $H_q$ as function of rank $q$ in MLLA
for different energies $Y$
for fixed $Q_c=Q_0$ 
(representing hadrons). With increasing $Y$ the moments approach the 
asymptotic behaviour 
 $H_q \simeq 1/q^2$.
}
\label{hqmlla-q}
\end{figure} 

Next we consider the moment ratios $H_q$ for hadrons ($Q_c=Q_0)$ from low to
high energies in Fig. \ref{hqmlla-q}. At the low energies ($Y\leq 4$) we
observe again the pattern of rapid oscillations from one order $q$ to the
next as in DLA reflecting the threshold behaviour. Beyond the Poissonian
transition point ($Y \geq 6$) we enter a new regime with a larger
oscillation length. The first minimum is slowly rising from $q=5$ at $Y=6$
to $q=7$ at $Y=10$. At higher $q$ a maximum follows. This is expected from
the analytical results \cite{dreminosc,dg}.
Whereas the oscillation amplitude of $H_q$ rises with $q$ at low
energies, it decreases at the high energies. However, it is always much
below the asymptotic DLA limit for large $q$, only for small $q$ below the
first minimum one can see
a convergence to this
 limit in the considered energy range.

\begin{figure}[t!]
\begin{center}
\includegraphics[angle=-90,width=13cm]{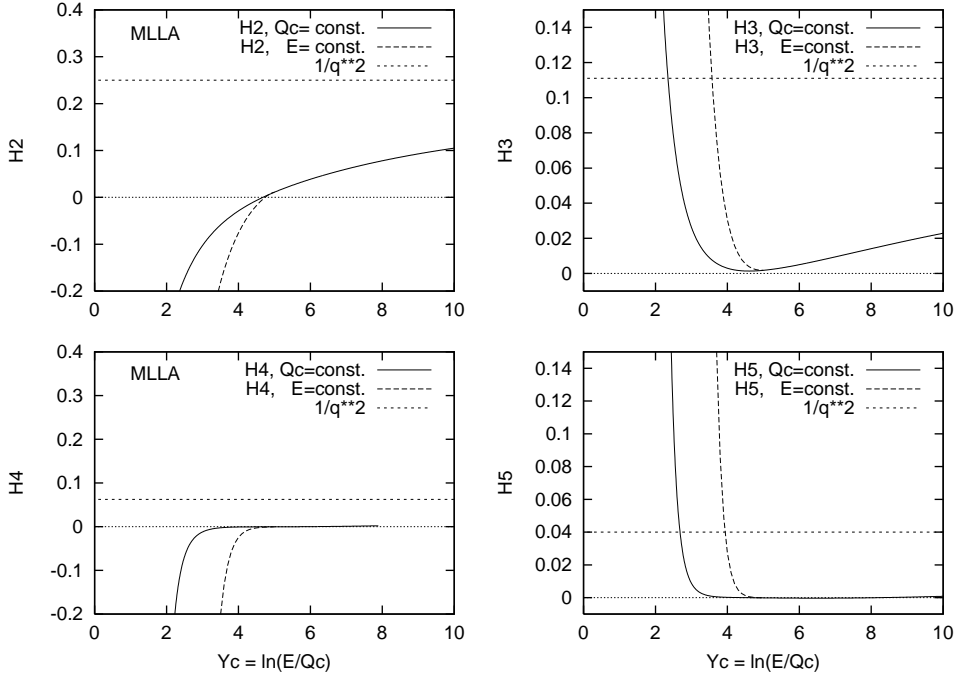}
\end{center}
\vspace{-0.3cm}
\caption{
Ratio of moments $H_q$ in MLLA vs energy for fixed $Q_c=Q_0$
(representing hadrons) and for fixed energy $Y=\ln(E/Q_0)=5$
 but variable $Q_c$ representing
jets at variable resolution $y_{cut}=(Q_c/2E)^2$. The asymptotic limits
$H_q\to 1/q^2$ are shown as well.
}
\label{hqmlla-yc}
\end{figure}

\begin{figure}[t!]
\begin{center}
\includegraphics[angle=-90,width=14cm]{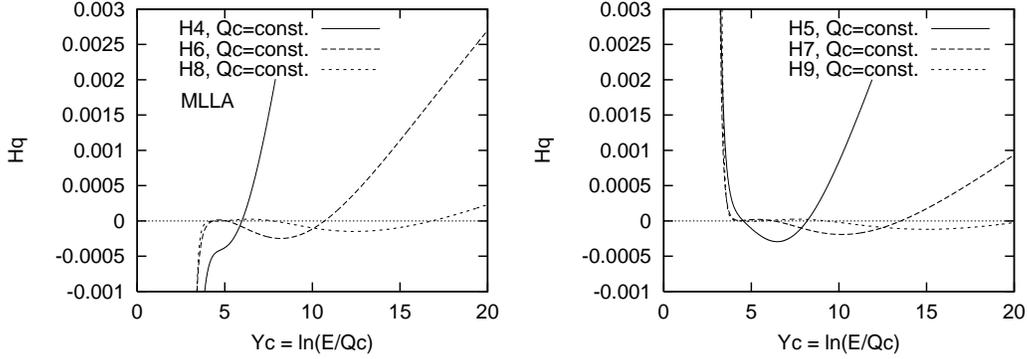}
\end{center}
\vspace{-5.0cm}
\caption{
Moment ratios $H_q$ in MLLA vs energy for fixed $Q_c=Q_0$
(representing hadrons) as in Fig. \ref{hqmlla-yc}.
The asymptotic limits of $H_q$ are above the upper bound of the plot.
}
\label{hq6-9mlla-yc}
\end{figure} 

The evolution of the $H_q$ moments with $Y_c$ is shown in Fig.
\ref{hqmlla-yc} for hadrons and jets. In comparison to the DLA results
of Fig. \ref{hqdla-yc}, we see at first a qualitatively similar behaviour,
but again the approach to the asymptotic limit is not visible for the higher
moments in the considered energy range.

More details are shown for hadrons in Fig. \ref{hq6-9mlla-yc}. 
New features emerge at high energies (above $Y\approx Y_P\approx 4$). 
For the higher moments ($q\geq 4$) secondary minima and/or
maxima appear with increasing energy at the level of ${\cal O}(10^{-4})$. 
These secondary oscillations
lead to a delay of the onset of the asymptotic behaviour. Below
$Y_c=20$ ($E\lsim 10^8$ GeV) the moments with $q\geq 6$ are below 1/10
of their DLA asymptotic value. 

The Poissonian transition point corresponds to $H_q=0$ for all $q>1$.
Indeed the moments show this behaviour in good approximation at 
the above $Y_P$. Beyond this point the positive short range correlations 
lead to a positive $H_2$, whereas below it the energy conservation
constraints lead to a negative cumulant. In general, the even  $H_q$ ratios
change sign because of the negative value at threshold
and the positive $1/q^2$ asymptotic limit.
In hadron phenomenology,
the positive correlations $H_2>0$ reflect the resonance production \cite{ddk}, 
in our
calculation it is the  gluon bremsstrahlung with small angle (above $Q_0$ cut-off) which lead to
the short range correlations, be it between ``hadrons''
(partons at scale $Q_0$) or between (mini) jets.  

As common features of DLA and MLLA, we note the splitting between 
hadron and jet moments at the same $Y_c$ which is a direct consequence of
the running coupling. Hadron and jet multiplicities coincide for $Q_c\to
Q_0$. 

\section{Monte Carlo simulation of parton cascade}
After the numerical solutions of the DLA and MLLA evolution equations for
single jets, we finally apply an MC event generator (ARIADNE \cite{ARIADNE}) 
at the parton level based on the same procedures as the above 
evolution equations: perturbative QCD
evolution with the coupling $\alpha_s(k_T)$ terminated by a
transverse momentum, $k_T$, cut-off and arbitrary parameters
$\Lambda$ and $Q_0>\Lambda$. The MC involves the coupled evolution 
of quarks and gluons in the cascade, the inclusion of large angle
radiation, the full first-order matrix element for $e^+e^-\to q\bar q g$,
and exact energy-momentum conservation. These features lead to a higher
accuracy than our MLLA calculation.\footnote{In MLLA, the inclusion of quark
jets is possible -- also, the first-order matrix element has been included in
the calculation of multiplicities \cite{lo}, however, 
energy-momentum conservation is fully realized in the MC approach and large
angle emission is more easily accessible.}
On the other hand, 
a simplification of the Monte Carlo is
the exclusive use of 1-loop calculations. This simplification could result
in deviations when very different energy scales are compared. 

In our adoption of the MC program, we set all quark masses to zero but kept
the masses for heavy quarks in the calculation of the number of flavours
$n_f$ for $\alpha_s$ at a given dipole mass.   
The results depend only on the two adjustable parameters 
$\Lambda$ and  $Q_0$; alternatively, 
we use the parameter $\lambda=\ln(Q_0/\Lambda)$. These parameters have been
determined from a fit to multiplicities. We refer to this modified
Monte Carlo as ``ARIADNE-D,'' where ``D'' stands for  ``Duality''.

The jet multiplicities are obtained from the final state partons using the
Durham algorithm, just as in the experimental analysis with (\ref{durham})
and (\ref{ycutlimits}). Hadron multiplicities are related to the parton
multiplicities, generated at scale $Q_0$, according to the LPHD prescription. 

\begin{figure}[t!]
\begin{center}
\includegraphics[angle=-90,width=13cm]{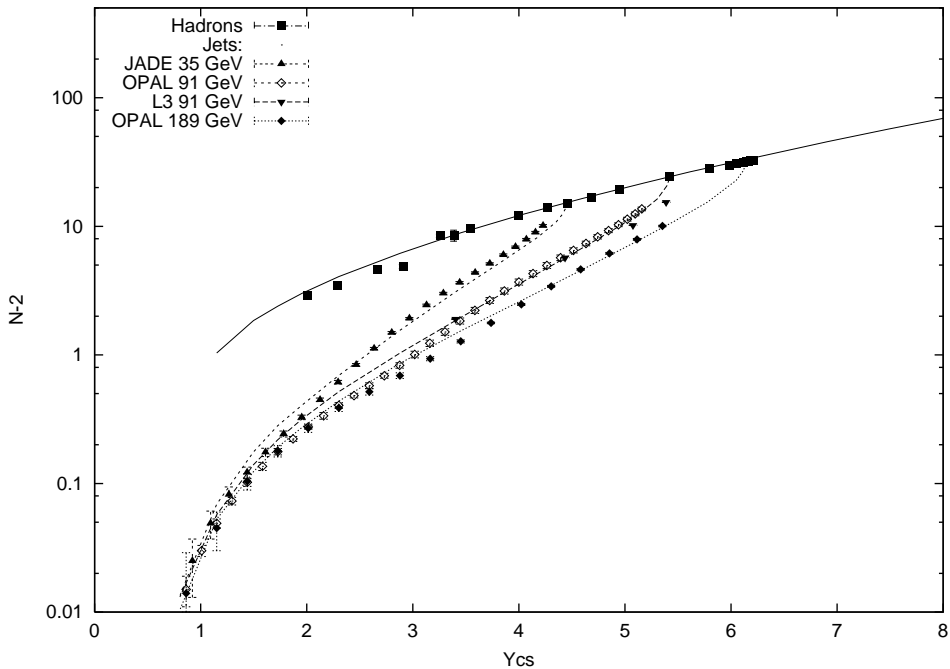} 
\end{center}
\vspace{-0.3cm}
\caption{
Multiplicity $N$ of hadrons 
(taken as $N_{ch}\times1.25$ using data
\cite{slacmult,argus,jade,tasso2,hrs,amy,%
alephgp,delphigp,l3gp,opalgp} and LEP averages from \cite{dg}) 
and multiplicity of jets \cite{opalpfeifen,L3osc} in $e^+e^-$ annihilation 
together with our Monte Carlo calculations as function of $Y_{cs}$,
 see (\ref{ycs}) ($Y_{cs} \sim \ln(Q/Q_c)$).
}
\label{multiplicity}
\end{figure}

This model will be compared with the experimental data. First we consider
the mean hadron and jet multiplicities $N$, as shown in Fig. \ref{multiplicity}. 
The jet multiplicities for energies $Q=35$, 91 and 189 GeV 
are obtained as  
functions of the cut-off $Q_c$ or of $y_{cut}=(Q_c/Q)^2$ -- 
in the figure they are plotted as a function of the variable 
\begin{equation}
Y_{cs}= -\frac{1}{2} \ln(y_{cut}+Q_0^2/Q^2) =\frac{1}{2}\ln(Q^2/(Q_c^2+Q_0^2)).
\label{ycs}
\end{equation}
with the additional scale $Q_0$. The MC parton multiplicity $N_{part}$ 
is shown at the cut-off
$Q_c=Q_0$ and should be compared at this scale with the total 
hadron multiplicity $N_{tot}$ according to the LPHD prescription,
in general $N_{tot}=K\times N_{part}$ (where $K$ is a constant); the observed charged multiplicity
is given by $N_{ch}=f_{ch} N_{tot}$ with the charged particle 
fraction $f_{ch}$.
These factors also depend on whether weak and electromagnetic decays 
of hadrons are included ($K^0,\Lambda,\ldots$). 
The factor $Kf_{ch}$ will be determined from the
fit. A simple choice is $K=1$ as for jets and $f_{ch}=2/3$ as obtained previously
\cite{lo}.\footnote{If 
proportionality of energy spectra of partons and
hadrons according to LPHD is required in the full energy range,
then the normalization has to be the same ($K=1$) 
because of energy conservation.}
For jets we compare parton and hadron results at the same cut-off and with
$K=1$. Note that
the choice of variable does not matter for the comparison of jet
properties (both MC and experimental 
data are obtained by the same procedure); but, 
using the variable $Y_{cs}$ in (\ref{ycs}), has the advantage that 
the jet data approach the
hadron data for $y_{cut}\to 0$ at a finite value of  $Y_{cs}$ 
in the figure according to (\ref{ycutlimits}).
In the region $Q_c=1-2$ GeV the dependence on $Q_0$ disappears and we are in
the perturbative QCD regime, governed by the single parameter $\Lambda$ 
\cite{woring}. 

\begin{table}[t]
\caption{Monte Carlo results on hadron and jet multiplicities 
for given values of $\Lambda$ (rows) and
$\lambda$ (columns). The
first half of the table consists of data with $y_{cut} = 0$ (hadrons), 
and the second half consists of
data with $y_{cut} = 2\times 10^{-5}$.}
\begin{center}
$
\begin{array}{c|ccllc}
\hline
\Lambda(\text{MeV}) \text{:}  & \lambda = 0.001 & \lambda = 0.01 & 
\lambda = 0.015 & \lambda = 0.05 \\
 \hline
250 & 35.8 & 31.7 & 30.6 & 26.5
\\
300 & 33.3 & 29.5 & 28.4 & 24.8
\\
400 & 29.6 & 26.3 & 25.4 & 22.0
\\
500 & 27.0 & 24.0 & 23.2 & 20.1
\\
\hline
\hline
250 & 12.4 & 12.2 & 12.1 & 11.8
\\
300 & 13.0 & 12.8 & 12.7 & 12.4
\\
400 & 13.9 & 13.6 & 13.5 & 13.0
\\
500 & 14.4 & 14.0 & 13.9 & 13.2
\\
\hline
  \end{array}
$
\end{center}
\label{tab:mult}
\end{table}

\begin{figure}[t!]
\begin{center}
\includegraphics[angle=-90,width=13cm]{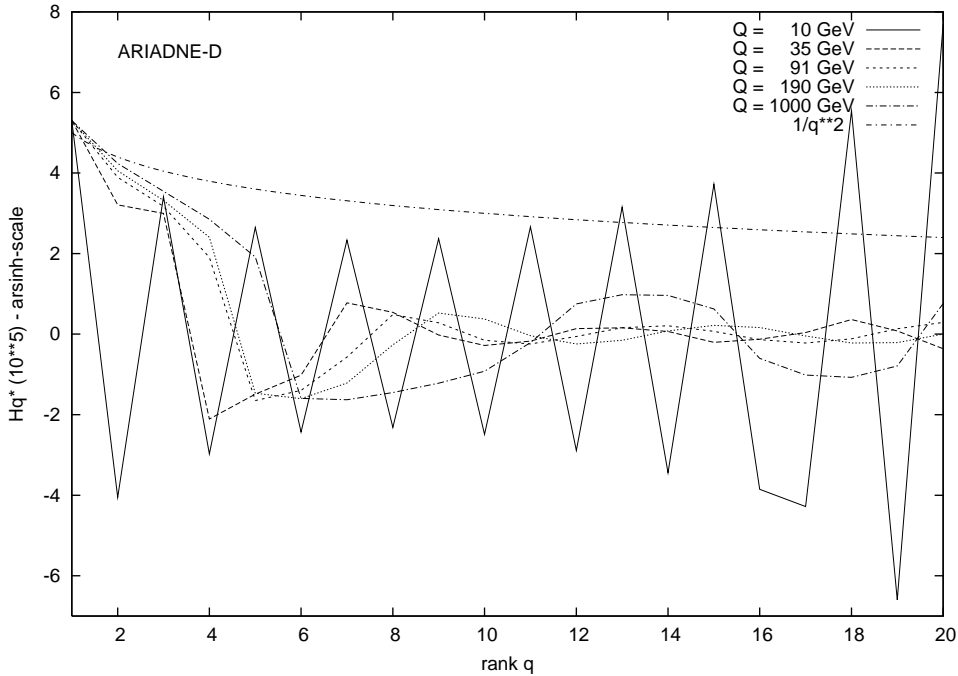} 
\end{center}
\vspace{-0.3cm}
\caption{
Ratios $H_q$ for a sequence of $cms$ energies $Q$ in $e^+e^-$ annihilation
as obtained from the Monte Carlo calculation, plotted as in Fig.
\ref{hqdla-q}.
}
\label{hq-ariadne-Q}
\end{figure}

\begin{table}[t!]
\caption{Ratios of multiplicity moments, $H_q$, for different cms energies, $Q$, in $e^+e^-$
 anihilation obtained from ARIADNE-D parton MC (the statistical errors of the MC results
  are generally smaller than the last digit given)}
\begin{center}
$
\begin{array}{r|ccccc}
\hline
q & 10 \ \text{GeV} & 35 \ \text{GeV} & 91 \ \text{GeV} & 190 \ \text{GeV} 
& 1000 \ \text{GeV}\\
 \hline
2 & -0.0565 & 0.00808 & 0.039 & 0.0569 & 0.0864
\\
3 & 0.0129 & 0.00495 & 0.0074 & 0.0104&0.0176
\\
4 & -0.0047 & -0.00064 & 0.00040 & 0.00126&0.00356
\\
5 & 0.0022 & -0.00015 & -0.00022 & -0.00014&0.00040
\\
6 & -0.0013 & -0.000051 & -0.00012 &-0.00020 &-0.00019
\\
7 & 0.0010 & 0.000028 & -0.000017 & -0.00008&-0.00021
\\
8 & -0.0010 & 0.000015 & 0.000013 & -0.000006&-0.00014
\\
9 & 0.0011 & -0.0000005 & 0.000006 &0.000015 &-0.00008
\\
10 & -0.0015 & -0.0000069 & -0.000003 &0.000009 &-0.00004
\\
11 & 0.0022 & -0.0000041 & -0.000005 & -0.0000009&-0.000005
\\
12 & -0.0038 & 0.0000032 & -0.000001 & -0.000005&0.000027
\\
13 & 0.0071 & 0.0000036 & 0.000003 & -0.000003&0.000047
\\
14 & -0.013 & 0.0000014 & 0.000004 & 0.000001&0.000045
\\
15 & 0.026 & -0.0000048 & 0.000001 & 0.000005&0.000019
\\
16 & -0.035 & -0.0000029 & -0.000003 & 0.000003&-0.000018
\\
17 & -0.09 & 0.0000008 & -0.000005 & -0.000001&-0.00005
\\
18 & 1.7 & 0.0000093 & -0.000002 & -0.000005&-0.00005
\\
19 & -19 & 0.0000018 & 0.000003 & -0.000005&-0.00003
\\
20 & 250 & -0.0000093 & 0.000007 & 0.000001 &0.00002
\\
\hline
  \end{array}
$
\end{center}
\label{tab:hq}
\end{table}

\begin{figure}[t!]
\begin{center}
\mbox{\epsfig{file=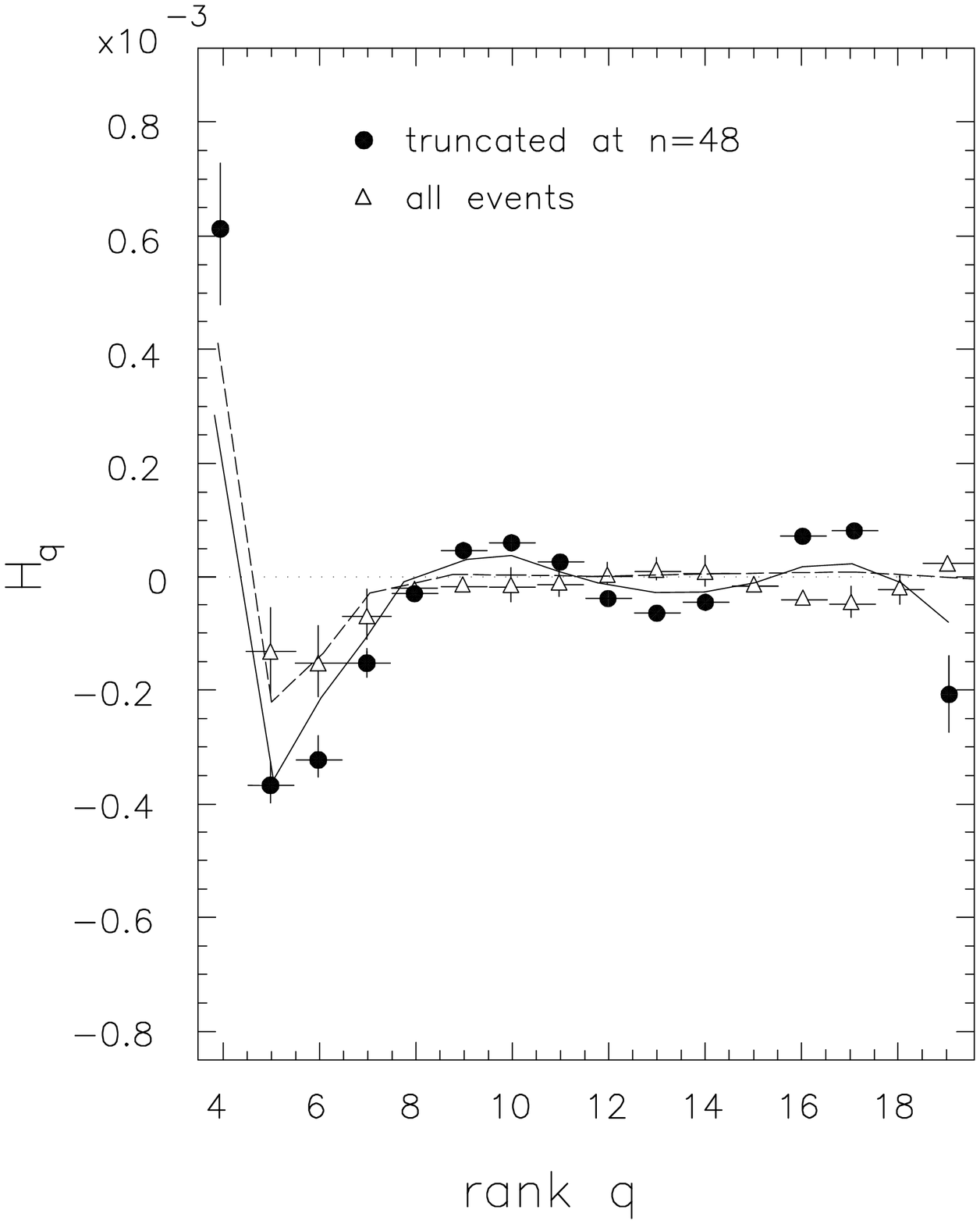,width=8.cm,bbllx=1.cm,bblly=5.5cm,%
bburx=19.cm,bbury=27.5cm}}
\vspace{-0.5cm}
\mbox{\epsfig{file=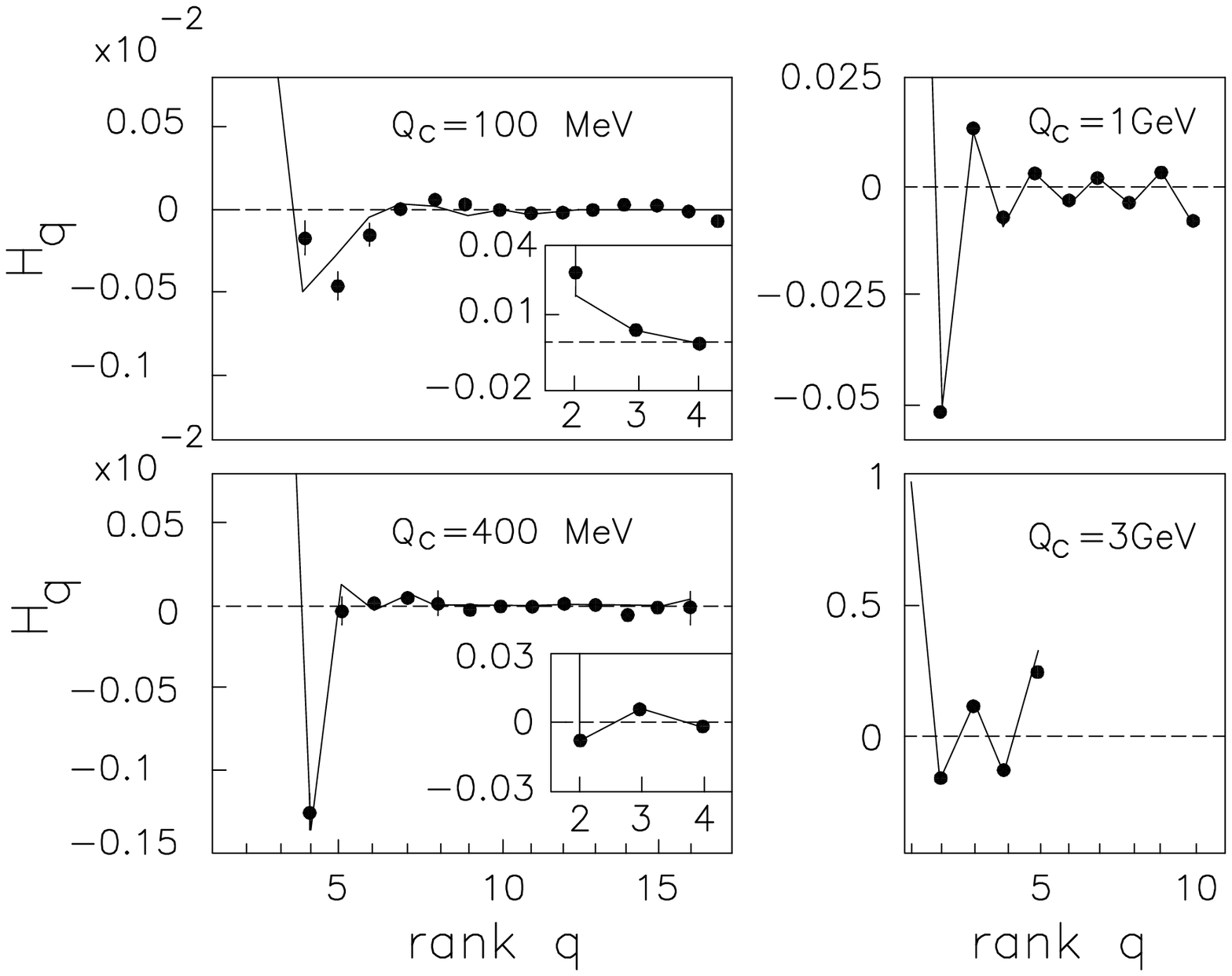,width=10cm,clip=,bbllx=2.cm,bblly=11.7cm,%
bburx=20.cm,bbury=26.5cm}}
\end{center}
\vspace{-0.5cm}
\caption{
(a) Ratio $H_q$ of multiplicity moments 
obtained by L3
Collaboration \protect\cite{L3osc,mangeol}
with truncation of multiplicity at $n_{ch}=48$ 
compared with ARIADNE-D MC (truncation at $n=65$) 
and both without truncation; (b)
Ratios $H_q$ for jets at different cut-off $Q_c$ 
 \protect\cite{L3osc} compared with our MC results.
}
\label{L3oscillations}
\end{figure}

In order to determine the parameters of the model
we compared the MC results with the multiplicity data at LEP-1. 
We noted that we could not get 
good agreement with the jet data from OPAL \cite{opalpfeifen} 
over the entire kinematic range of $10^{-5} <y_{cut} < 0.5$. 
Since we were interested in studying the
transition between jets and hadrons, we chose our parameters so that they
gave a particularly 
good description of the low $y_{cut}$ regime and the hadron
multiplicity as well as the very large $y_{cut}$ region.
In Table \ref{tab:mult} we illustrate the dependence of the
hadron and jet multiplicity on the parameters. 
We require a good fit of
the jet multiplicity of $N = 14.1\pm 0.4$ 
for $y_{cut}=2\times 10^{-5}$ \cite{opalpfeifen}
corresponding to $Q_c=0.4$ GeV. At this scale the particles from 
non-hadronic decays are recombined into the primary hadrons and this
ambiguity is removed.
Also we considered
the mean charged multiplicity at LEP-1 of $N_{ch}=21.2\pm 0.3$
\cite{dg} (this number includes particles from $K^0$ and $\Lambda$ decays). 
Solutions can be found close to the simple case
$f_{ch}K= 2/3$.
We do not require this condition though and choose the parameters 
\begin{equation}
\Lambda=400\ \text{MeV}, \quad \lambda=0.01, \quad (Kf_{ch})^{-1}=1.25;
\label{parameters}
\end{equation} 
they are not very different from the previous study for hadrons alone
 ($\Lambda=200$ MeV, $\lambda=0.015$ and $(Kf_{ch})^{-1}=1.5$
\cite{low}) which optimized the description of the hadronic correlations
alone.

With our choice (\ref{parameters}) 
a reasonable description of jet multiplicities 
for small and large $y_{cut}$
 was obtained, not only at LEP-1, but also at
the higher and lower energies of 35 and 189 GeV, 
as can be seen in Fig. \ref{multiplicity}. Furthermore we get a good
description of the trend of
hadron multiplicities as function of
$cms$ energy in the full range $Q=3\ldots 200$ GeV. 
The Monte Carlo was run with 250,000 events at each energy. 

 As in the analytic calculations,
the multiplicity in the MC approach applied here 
would diverge for $Q_0\to \Lambda$. Now,
Fig. \ref{multiplicity} shows that with the same cut-off $Q_0$ 
taken at all energies the correct
energy dependence of the hadron multiplicity is obtained.

At every $cms$ energy in Fig. \ref{multiplicity}, the MC fails to duplicate the
 intermediate $y_{cut}$ region with MC results falling above the experimental data.
We find that
the structure we are unable to reproduce occurs at the fixed b mass scale 
$Q_c\sim 5$ GeV at all energies. In our Monte Carlo we use
only light quarks so as to avoid complicated decay processes. 
It is plausible to assume, then, that the 'hump' structure at 5 GeV 
comes from b quark production. 
This conjecture is further supported by 
 the enhancement of the effect seen at the $Z$ energy of 91 GeV
as expected from the enhanced  b quark production by
neutral current interactions of the $Z$.

The results in Fig. \ref{multiplicity} look similar to the analytic
solutions in Figs. \ref{hqdla-nf} and \ref{hqmlla-nf} 
and demonstrate again the importance of the
running coupling which leads to the scale ($Q$) dependence 
of multiplicity at fixed $Y_{cs}$ or $Q/Q_c$; in the threshold region (low
$Y_{cs}$) the difference between hadron and jet multiplicities amounts to a
an order of magnitude.

\begin{figure}[h!]
\begin{center}
\includegraphics[angle=-90,width=10cm]{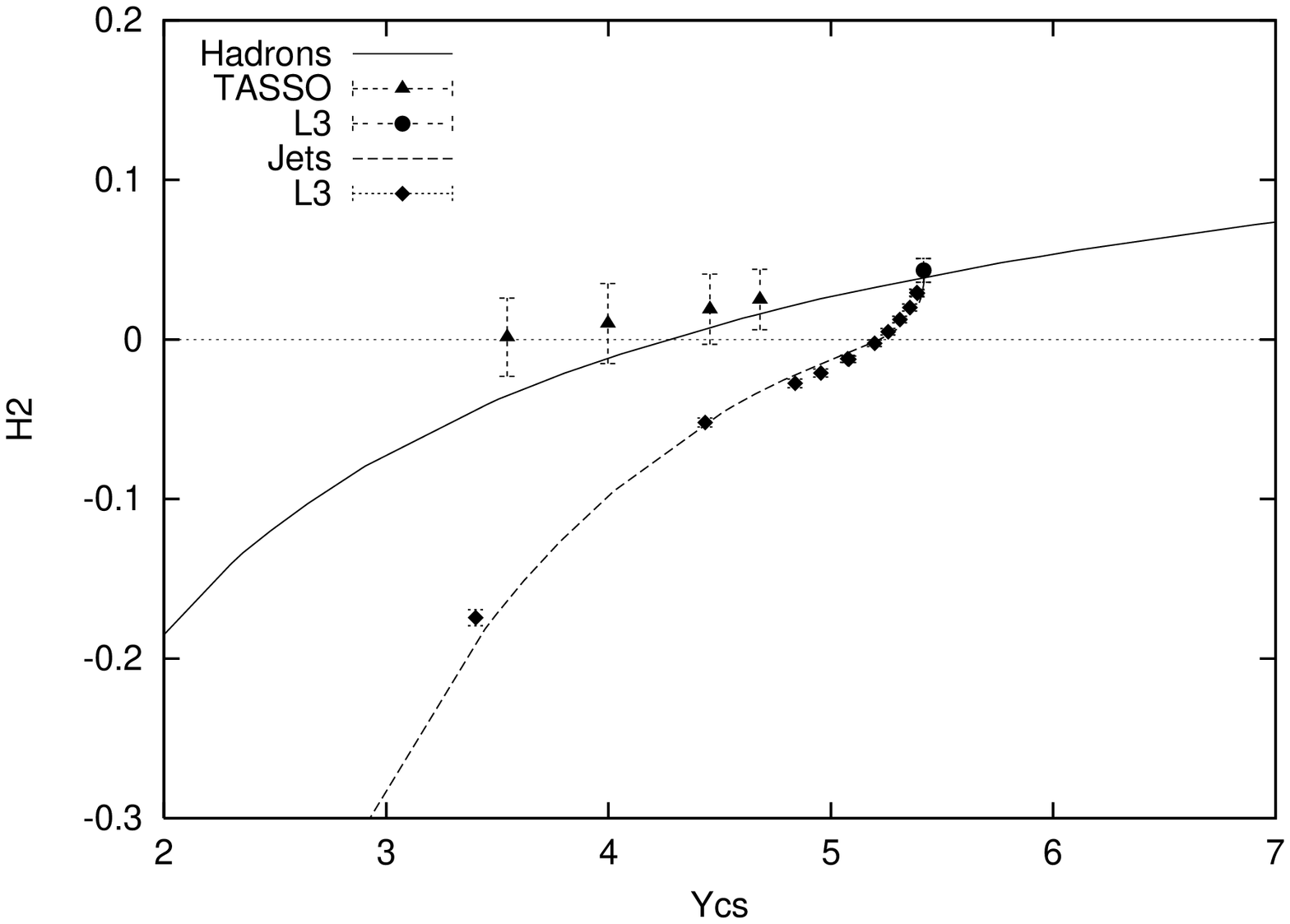}\\
\includegraphics[angle=-90,width=10cm]{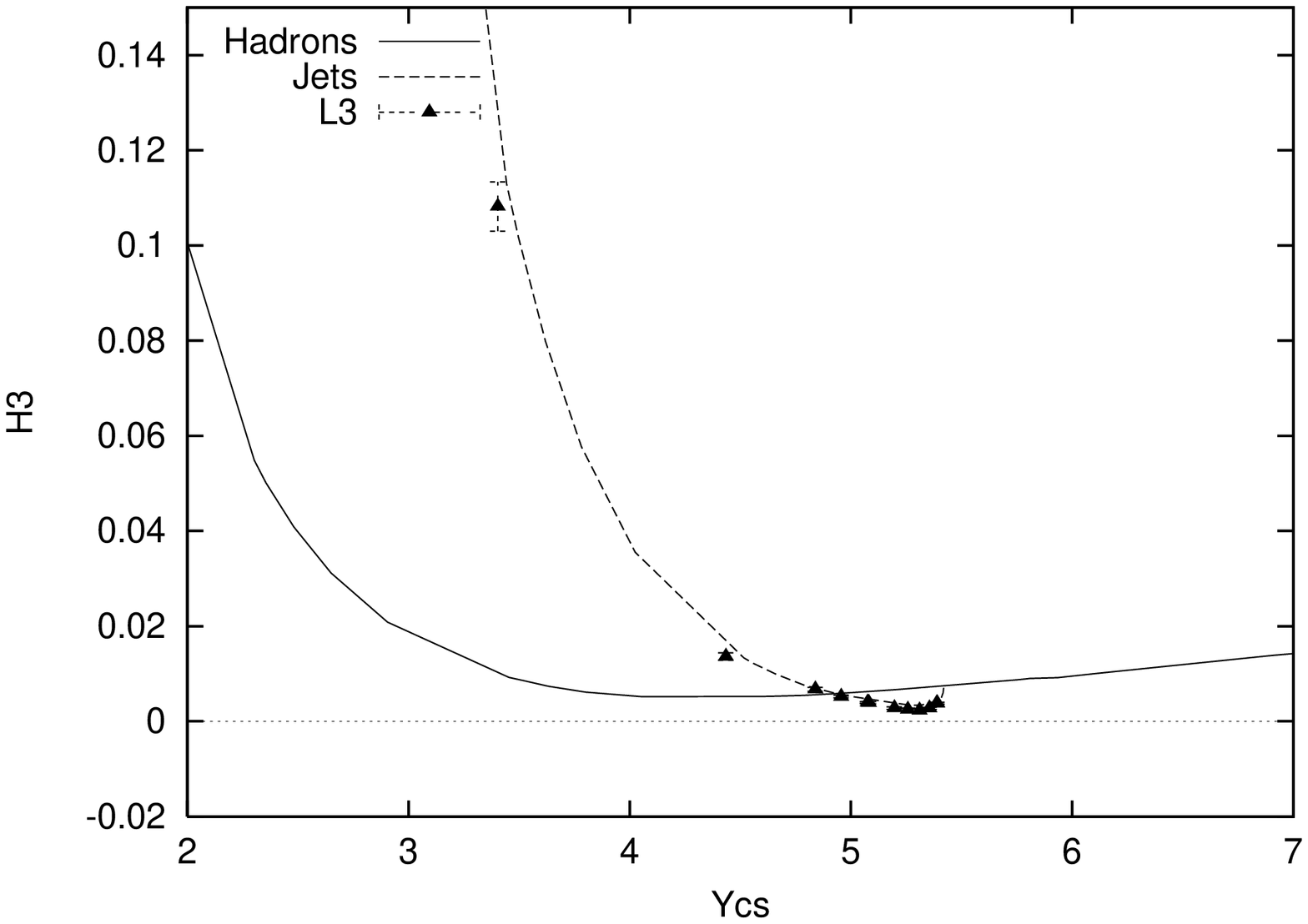}\\
\end{center}
\vspace{-0.3cm}
\caption{    
Ratio of moments $H_q$ for hadrons \cite{tasso} and jets
\cite{L3osc,mangeol} in comparison with ARIADNE-D Monte Carlo.}
\label{hq2-3exp}
\end{figure} 

Next, we turn to the energy evolution of the $H_q$ moments 
for hadrons (i.e. partons
at scale $Q_0$) from low to high energies (10-1000 GeV).
This evolution is displayed in Fig. \ref{hq-ariadne-Q} and looks similar to
the result from MLLA in Fig. \ref{hqmlla-q}: at the low energy, below the
Poissonian point, there are the rapid oscillations reflecting the threshold 
behaviour, above that point we observe the oscillations with increasing length 
for increasing energy. The first minimum occurs at $q_{min}\approx5$ at 90 GeV 
 and increases to $q_{min}\approx7$ at 1000 GeV. 
This increase is similar but a bit stronger 
than in the asymptotic formula (\ref{qmin}),
which would predict an increase by one unit of $q_{min}$ instead of two.
These calculations have been carried out with a sample of 4M events.
The statistical errors are determined from the fluctuations of 4 sub-samples
and are found to be significantly smaller
than the oscillation amplitudes at their maximum. The numerical results are
also given in Table \ref{tab:hq}.

\begin{figure}[t!]
\begin{center}
\includegraphics[angle=-90,width=10cm]{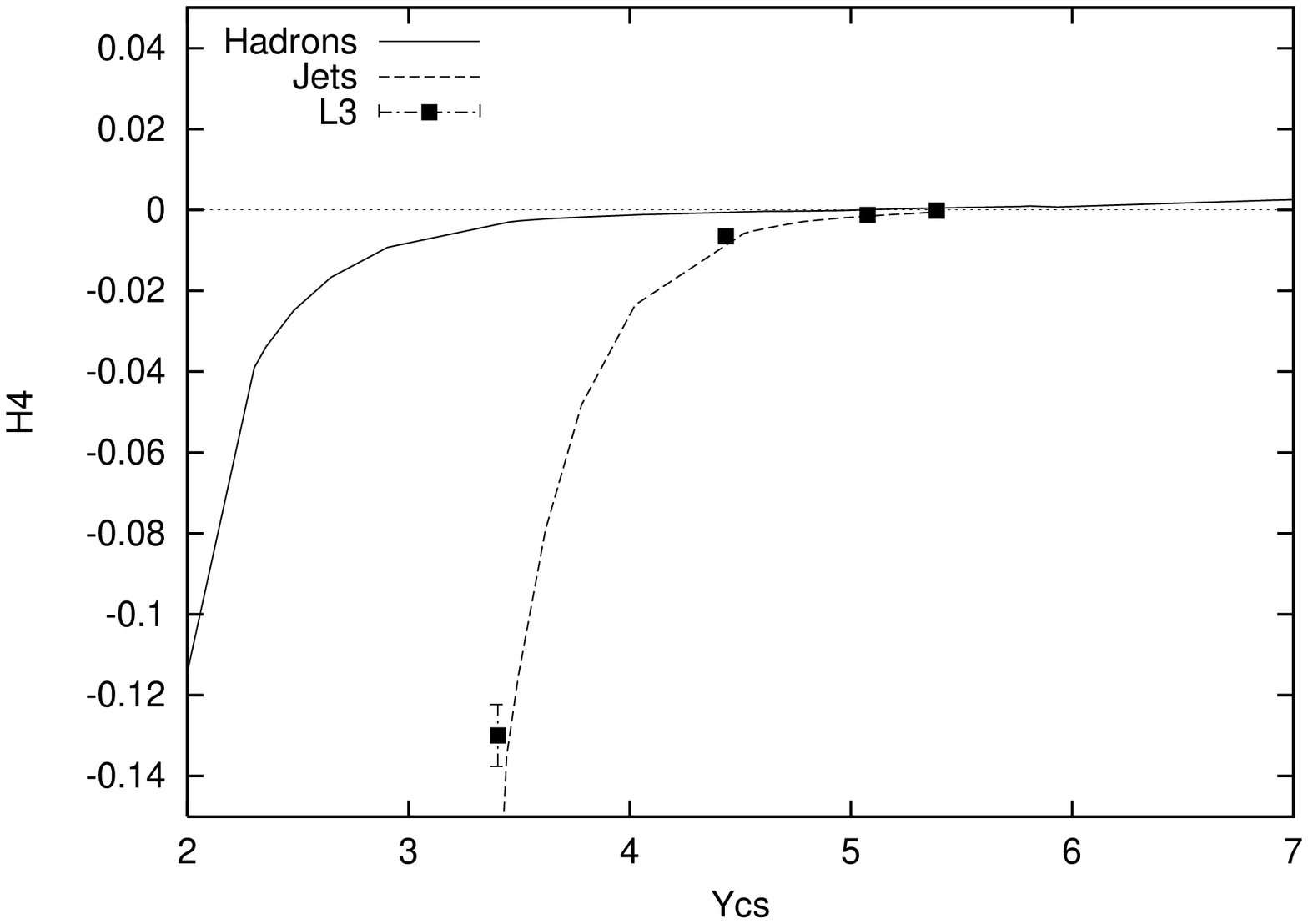}\\
\includegraphics[angle=-90,width=10cm]{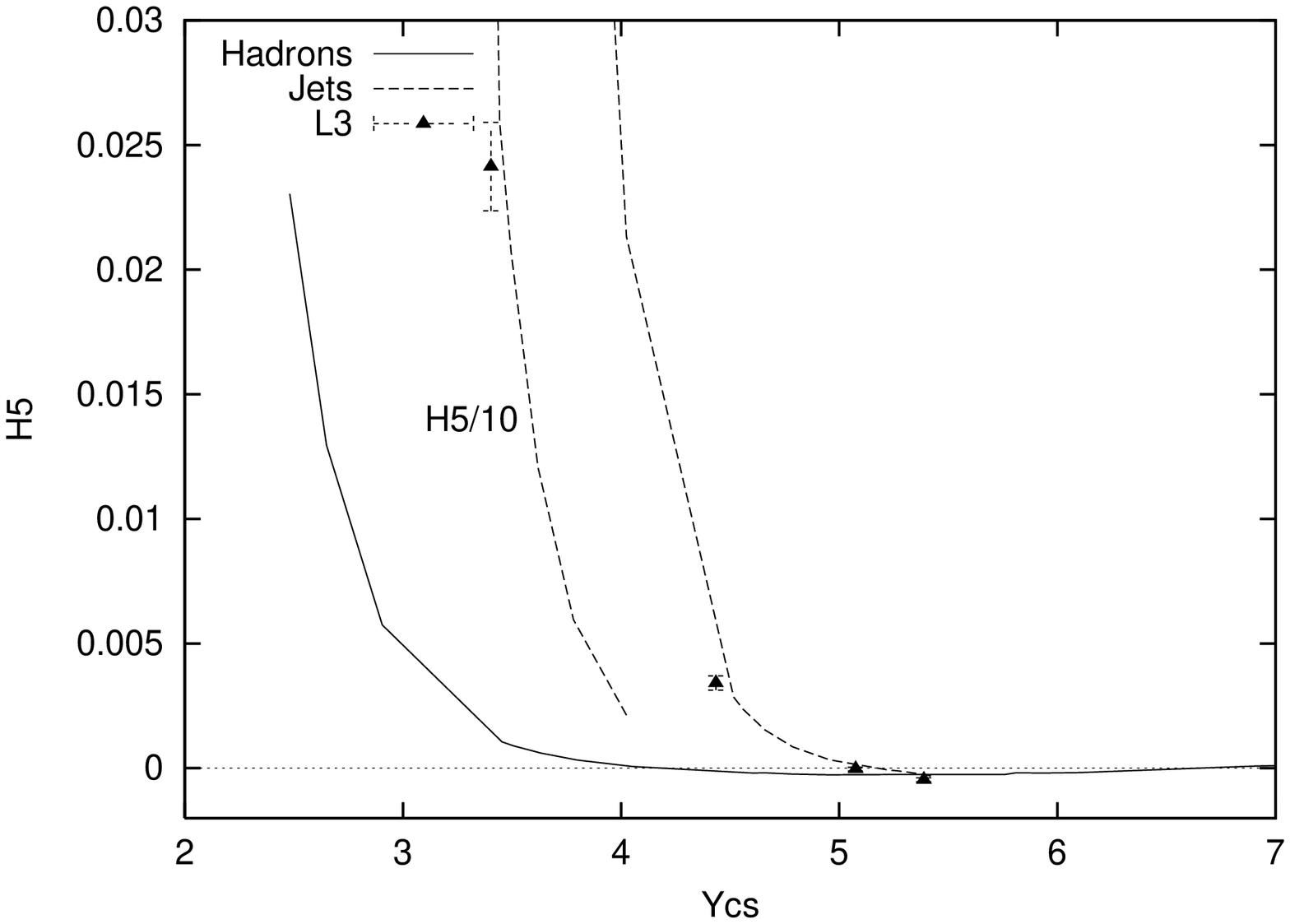}
\end{center}
\vspace{-0.3cm}
\caption{
Ratio of moments $H_q$ for hadrons and jets as in Fig. \ref{hq2-3exp}
}
\label{hq4-5exp}
\end{figure} 

\begin{figure}[t!]
\begin{center}
\includegraphics[angle=-90,width=10cm]{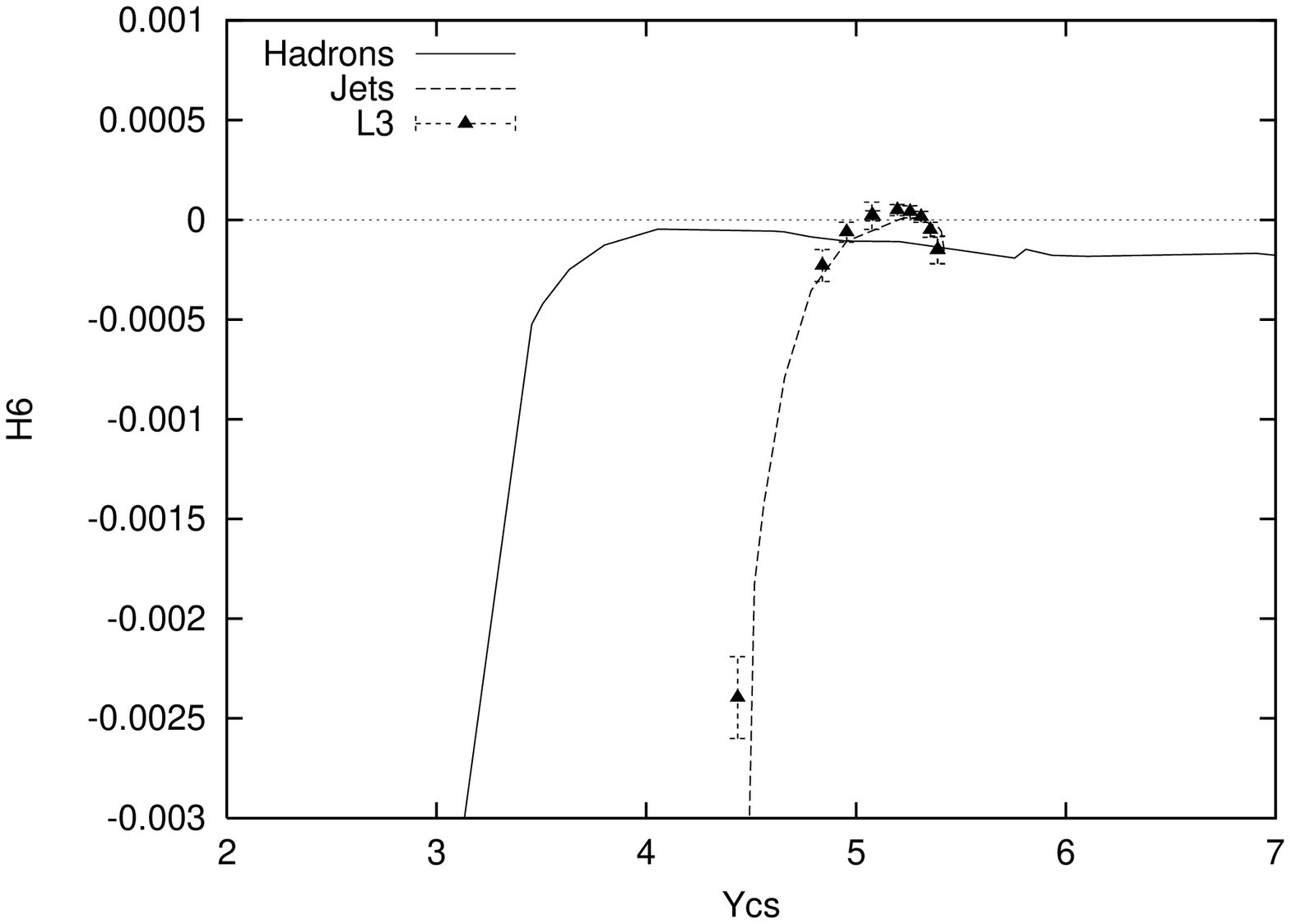}\\
\includegraphics[angle=-90,width=10cm]{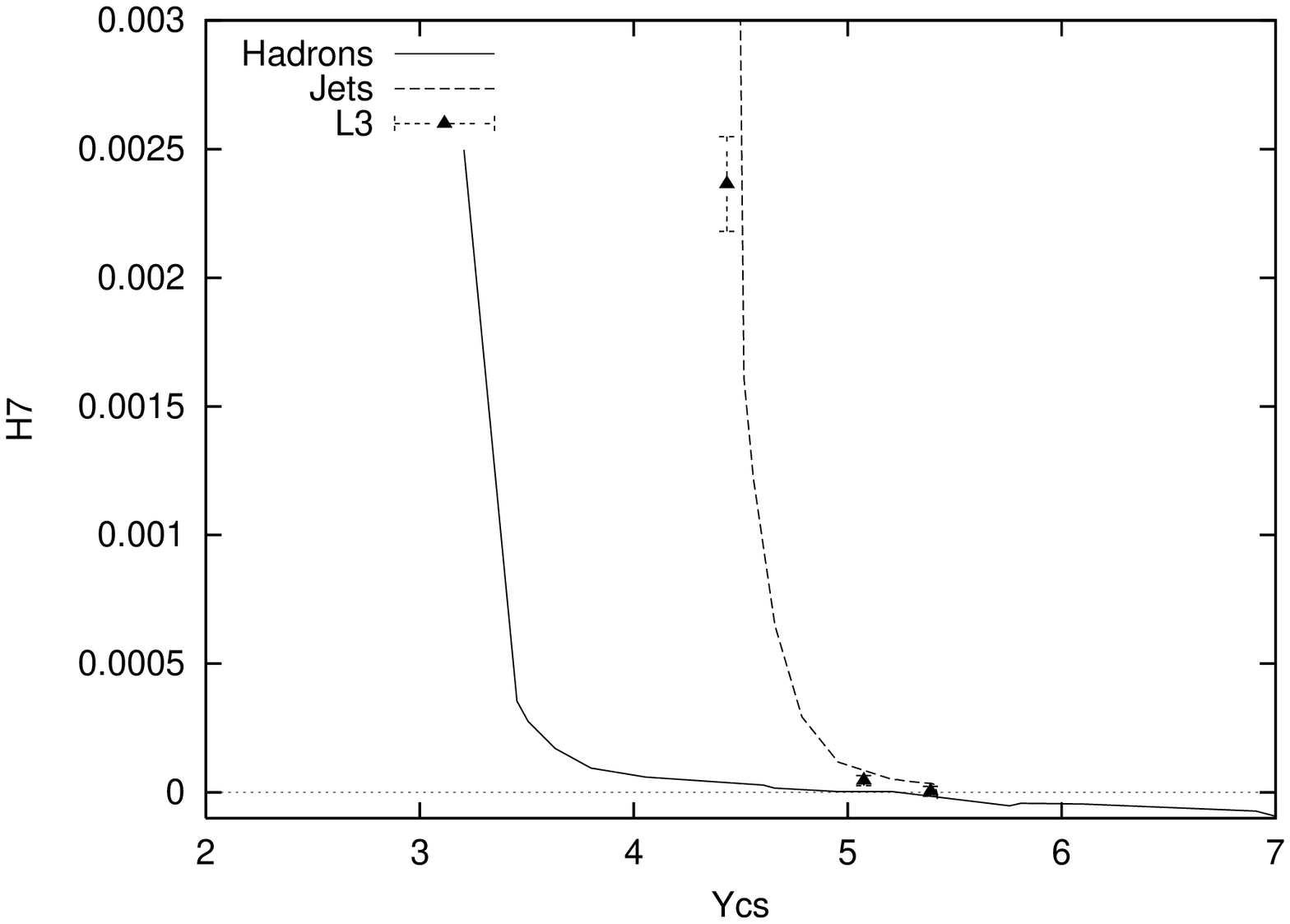}
\end{center}
\vspace{-0.3cm}
\caption{
Ratio of moments $H_q$ for hadrons and jets as in Fig. \ref{hq2-3exp}
}
\label{hq6-7exp}
\end{figure}

In Fig. \ref{L3oscillations}, we compare the ratios, $H_q$, obtained from our
calculations with the measurements by L3,
both for hadrons and for jets with variable
(Durham) cut-off $Q_c$. For hadrons we assume that the multiplicity 
distribution of charged particles has the same normalized moments as the
distribution of all particles. This is the case if the same conversion
factor $f_{ch}$ applies for all multiplicities $n$.

For hadrons, a measurement of the full
multiplicity distribution \cite{mangeol} which extends up to $n_{ch}=62$
and one with truncation \cite{L3osc} 
at the multiplicity
$n_{ch}=48$ are presented. The truncation can be seen to considerably increase
 the oscillation amplitude. We compare with the Monte Carlo results 
again with and
without truncation. The truncation was at multiplicity $n_t=63$ to obtain the
same event fraction of 0.005\% for multiplicities $n>n_t$ as in the
experimental data. Our results reproduce the data rather well but for the
truncated moments the high $q$ oscillations are a bit weaker.

We mention here another subtlety of the experimental analysis. The L3 data
in Fig.  \ref{L3oscillations}  refer to data without $K^0$ and $\Lambda$
decays, whereas the multiplicities in Fig. \ref{multiplicity} include them.
However, the $H_q$ moments do not depend on this difference, except to 
within a few per
cent. The same is true  for the truncated moments, as long as the removed event 
fraction is the same \cite{mangeol}. This appears plausible from the point
of view of KNO scaling.

We note that the moments presented by the SLD Collaboration \cite{slacosc}
have been determined from the multiplicity range 
$6 \leq n_{ch}\leq 54$ and they look very similar to the truncated L3
moments. We stress that from our point of view the
untruncated moments are of primary relevance; only these obey the evolution
equations discussed above and have simple asymptotic properties.
Beyond the first minimum they are very small at LEP energies, of the order
of $10^{-5}$.

Finally, in the same Fig. \ref{L3oscillations} 
we study the evolution of the $H_q$ moments with cut-off $Q_c$ or 
$y_{cut}=(Q_c/Q)^2$. The MC calculations reproduce well the change 
from the rapid oscillations with large amplitude at large cut-off $Q_c/Q$
 to the oscillations with larger oscillation length and smaller
amplitudes in the small $Q_c/Q$ region close to the fully resolved hadronic
final state.

In the subsequent Figs. \ref{hq2-3exp}
-\ref{hq6-7exp}
we show these ratios, $H_q$, 
separately as function of $Y_{cs}$, i.e. of $Q/Q_c$.  
Included are further data at intermediate $y_{cut}$ values from
\cite{mangeol} together with the corresponding MC predictions and also the
predictions for the hadronic final state, i.e. at $Q_c=Q_0$.
In these calculations 2.5 M events have been generated in the MC. 
For all orders
$q$, the calculation follows the experimental data closely.
Again, there is the same qualitative difference between the hadron and jet data
as observed before for the DLA and MLLA calculations in Figs. \ref{hqdla-yc}
and \ref{hqmlla-yc}. 

For the ratio, $H_2$,  we also
compare with results we deduced from TASSO moments \cite{tasso}. The data
clearly confirm the splitting of the moments at different scales, as
expected from the running coupling. At small $Q/Q_c$, both for hadrons and
for jets, the distribution is narrower than Poisson ($H_2<0$), while at 
higher values
of this ratio the distribution gets broader ($H_2>0$). The positive $H_q$
is expected from the short range correlations, either 
from gluon bremsstrahlung at
small angles for partons or from resonance decays for hadrons.
At the point with
$H_2=0$ the higher moments are also close to zero corresponding to a
Poissonian transition point, both for jets and hadrons: 
\begin{equation}
\text{Poissonian transition point:} 
\qquad 
\begin{cases} Q\approx 30\  \text{GeV}   & \text{hadrons}\\
              Q=91,\ Q_c\approx 0.3\ \text{GeV} & \text{jets}.
\end{cases}
\label{poissonptdat}
\end{equation}
It is interesting to note that the Poissonian distribution has actually been
successfully fit to their data at 29 GeV by the HRS Collaboration
\cite{hrs}.
 
We observe again a close similarity between MC and MLLA results, 
although the deviations from a 
Poissonian ($H_3>0$ at minimum) appear a bit larger in the MC.
 
\section{Conclusions}
We have studied the multiplicity moments, in particular the $H_q$ ratios
for the QCD parton cascade in three different approximations: DLA, MLLA and
parton Monte Carlo. The DLA predicts the very asymptotic behaviour. The
asymptotic MLLA corrections predict the existence of $H_q$ oscillations,
but do not explain any $y_{cut}$ dependence. The
more precise calculations of the full MLLA solution of the evolution
equations and the Monte Carlo method, 
which include the correct threshold
behaviour, show at the presently accessible energies the following new
features of the multiplicity moments:
\begin{itemize} 
\item[a)] There is a particular energy scale, $Y_P$,
 for hadrons or a particular $y_{cut}$ for jets where the multiplicity
distribution is approximately Poissonian ($H_q\approx0,\ F_q\approx1$ for
$q>1$).
\item[b)] In a region of smaller
energies there is a rapid oscillation of $H_q$, reflecting the threshold
behaviour; at higher energies  the
oscillation length increases with energy as predicted qualitatively by
asymptotic MLLA, but the oscillation amplitude is much smaller. 
In fact, for hadrons at LEP energies we predict $|H_q|\lsim 10^{-5}$
beyond the first minimum at $q\approx 5$. Also, there are 
secondary extrema of $H_q$ as a function of energy for a given rank $q$.
The approach to the asymptotic DLA region is very slow if visible at all.
\item[c)] For fixed resolution parameter $y_{cut}$ the moments of jets and
hadrons are rather different as a consequence of the running coupling.
\end{itemize}

We conclude that the perturbative approach yields good results not only for
single particle phenomena, such as single inclusive spectra and
mean multiplicities, as envisaged originally in the LPHD approach, 
but also for correlations of high order. 
Above the Poissonian transition point, the short range correlations
initiated by resonance decays for hadrons and gluon Brems\-strahlung for
partons  
become increasingly important and lead to the broadening of the multiplicity
distribution of both hadrons and jets ($H_2>0$). So there is a duality also
between hadronic and partonic correlations.

Moreover, the absolute normalization, originally considered as free
parameter, can be studied as function of resolution $y_{cut}$ and a unified
description of hadronic and jet phenomena is possible, using the variable
$Y_{cs}$ in (\ref{ycs}) with the shifted $y_{cut}$ prescription, both 
for multiplicities and correlations. 

Both the predicted energy evolution of hadronic correlations and the  
$y_{cut}$ 
dependence of the jet correlations follow the experimental results closely,
with the exception of jets near the $b$ quark threshold.
In the jet regime $Q_c\gsim 1-2$ GeV, the effect of the hadronic scale $Q_0$ 
becomes
negligible and we are in the domain of perturbative QCD with the single
parameter $\Lambda$ only; $Q_c$ is then an external parameter to be chosen
by the experimenter. At smaller scales, the
hadronic scale $Q_0$ becomes important, and the results become model dependent. 
The transition region from jets to hadrons for $y_{cut}\to 0$, 
which shows strong variations of all moments, is very well
reproduced by the parton MC with low cut-off $k_T\geq Q_0>\Lambda$. In this
region the coupling becomes large, i.e. $\alpha_s \gsim 1$. 
An extension of the calculation into this kinematic region is not justified
a priori, but, as seen from the
analytical results, the  
convergence of the perturbation theory is not in danger, 
and the solutions can be obtained from the 
evolution equations or from the MC. This successful description
can be viewed at least as a good effective parametrization of the soft
transition region.

With increasing order, accurate calculations are required. 
So far, this is only accessible by numerical methods. 
Further improvements are possible by including heavy quarks in the parton
evolution and by including the 2-loop results.

The good description of the data by the MC model with 
normalization $K\approx 1$ 
implies that in the considered applications, one parton 
counts about one hadron and the
 hadron final state is well
represented by a parton final state of the same multiplicity 
at resolution scale $Q_0\gsim \Lambda$. 

\section*{Acknowledgement}
We thank Valery Khoze for the interesting 
discussions and the useful comments on the manuscript.


  
\end{document}